\documentclass[twocolumn, secnumarabic,amssymb, nofootinbib, notitlepage, aps, prd]{revtex4-1}
\usepackage{dcolumn}   
\usepackage{bm, bbm, bbding}
\usepackage{amssymb,amsmath,amsfonts,mathrsfs,graphicx,xcolor,units,yfonts}
\usepackage{txfonts,dsfont,extarrows}
\usepackage[toc,page,title,titletoc,header]{appendix}
\usepackage{setspace, framed, hyperref}
\usepackage[novbox]{pdfsync}

\hypersetup{
  colorlinks=true,        
  linkcolor=blue,         
  citecolor=magenta,       
  urlcolor=black  }

\begin{document}

\title{Apparent horizon and gravitational thermodynamics of the Universe: Solutions to the temperature and entropy confusions, and extensions to modified gravity}

\author{David Wenjie Tian}%
\email[]{wtian@mun.ca}
\affiliation{Faculty of Science,  Memorial University, St. John's, Newfoundland, Canada, A1C 5S7}
\author{Ivan Booth}%
\email[]{ibooth@mun.ca}
\affiliation{Department of Mathematics and Statistics, Memorial University, St. John's,  Newfoundland, Canada, A1C 5S7}
\begin{abstract}

The thermodynamics of the Universe is restudied by requiring its compatibility with the holographic-style gravitational equations which govern the dynamics of both the cosmological apparent horizon and the entire  Universe, and possible solutions are proposed to the existent confusions regarding the apparent-horizon temperature and the cosmic entropy evolution. We start from  the generic Lambda Cold Dark Matter  ($\Lambda$CDM) cosmology of general relativity (GR) to establish a framework for the gravitational thermodynamics. The Cai--Kim Clausius equation $\delta Q=T_{\text{A}}dS_{\text{A}}=-dE_{\text{A}}=-A_{\text{A}}\bm\psi_t$ for the isochoric process of an instantaneous apparent horizon indicates that, the Universe and its horizon entropies encode the \emph{positive heat out}  thermodynamic sign convention, which encourages us to adjust the traditional  positive-heat-in Gibbs equation into the  positive-heat-out version $dE_m=-T_mdS_m-P_mdV$. It turns out that  the standard and the generalized second laws (GSLs) of nondecreasing entropies are always respected by the event-horizon  system as long as the expanding Universe is dominated by nonexotic matter $-1\leq w_m\leq 1$, while for the apparent-horizon simple open system the two second laws hold  if $-1\leq w_m<-1/3$; also,  the artificial local equilibrium assumption is abandoned  in the GSL. All constraints regarding entropy evolution are expressed by the equation of state parameter, which show that from a thermodynamic perspective  the phantom dark energy is less favored than the cosmological constant and the quintessence. Finally, the whole framework is extended from GR and $\Lambda$CDM to modified gravities with field equations  $R_{\mu\nu}-Rg_{\mu\nu}/2=8\pi G_{\text{eff}}  T_{\mu\nu}^{\text{(eff)}}$. Furthermore, this paper argues that the Cai--Kim temperature is more suitable than Hayward, both temperatures are independent of the inner or outer trappedness of the apparent horizon, and the Bekenstein--Hawking and Wald entropies cannot unconditionally apply to the event and particle horizons.\\

\noindent PACS numbers: \;\;04.20.Cv \,,\, 04.50.Kd \,,\,  98.80.Jk
\end{abstract}

\maketitle

\section{Introduction}\label{section Introduction}


The thermodynamics of the   Universe is quite an interesting problem and has attracted a lot of discussion. Pioneering work dates back to the investigations of cosmic entropy evolutions for the spatially flat  de Sitter Universe  \cite{de Sitter early 1} dominated by a positive cosmological constant, while recent studies have covered both the first and second laws of thermodynamics for the Friedmann-Robertson-Walker (FRW) Universe with a  generic spatial curvature.

Recent interest on the first law of thermodynamics for the Universe was initiated by Cai and Kim's  derivation of the Friedmann equations from a thermodynamic approach \cite{Cai I}: this is actually a continuation of Jacobson's work to recover Einstein's equation from the equilibrium Clausius relation on local Rindler horizons \cite{Jacobson 1995}, and also a part of the  effort to seek   the connections between thermodynamics and gravity \cite{Thermodynamical aspects of gravity} following the discovery of black hole thermodynamics \cite{Bekenstein-Hawking entropy}.
For general relativity (GR), Gauss-Bonnet and Lovelock gravities, Akbar and Cai reversed the formulation in  \cite{Cai I} by rewriting the Friedmann equations into the heat balance equation and  the unified first law of thermodynamics   at the cosmological apparent horizon \cite{Cai V}.
The method of \cite{Cai V} was soon generalized to  other theories of gravity to construct the effective total energy differentials by the corresponding modified Friedmann equations, such as the  scalar-tensor gravity in \cite{Negative T=kappa/2pi I},  $f(R)$ gravity
in \cite{Cai IV}, braneworld scenarios in \cite{Inverse brane world, Inverse brane world T absolute}, generic $f(R,\phi,\nabla_\alpha\phi \nabla^\alpha \phi)$ gravity in \cite{scalar f R phi phi2}, and Horava-Lifshitz gravity in \cite{UFL Horava-Lifshitz}.  Also, at a more fundamental level, the generic field equations of $F(R ,\phi ,-\frac{1}{2}\nabla_\alpha\phi \nabla^\alpha \phi ,\mathcal{G})$  gravity are recast into the form of  Clausius relation in \cite{Clausius modified gravity}.


Besides the first laws on the construction of various energy-conservation and  heat-transfer equations, the entropy evolution of the Universe has also drawn plenty of attention.
However, the cosmic entropies are almost exclusively studied in the generalized
rather than the standard second laws \cite{causal boundary entropy, GSLT quintom, GSL Interacting holographic DE, EventH GSL fR flat,  2nd law modified gravity, GSL MG Akbar, GSL Gauss-Bonnet braneworld, GSL Horava-Lifshitz, EventH GSLT Gauss Bonnet, GSL f(T) gravity, GSL scalar tensor chameleon, GSLT f R interacting}. In fact, investigations via  the  traditional Gibbs
equation $dE_m=T_md\tilde{S}_m-P_mdV$ show that in GR and modified gravities, the evolution
of the physical entropy $\tilde{S}_m$ for the matter inside  the apparent
and the event horizons  departs dramatically from the desired nondecreasing behaviors; especially that  $\tilde{S}_m$ inside the future-pointed event horizon always decreases under the dominance of nonexotic matter above the phantom divide. Thus the generalized second law (GSL) has been employed, which adds up  $\tilde{S}_m$  with   the geometrically defined entropy of the cosmological causal boundaries and anticipates the total entropy to be nondecreasing so that  the standard second law could be rescued. For example, GSL has been studied in \cite{causal boundary entropy} for a flat Universe with  multiple entropy  sources (thermal, geometric, quantum etc.) by the entropy ansatz $S=|H|^\alpha$ ($\alpha>-3$),   in \cite{GSLT quintom} for the event-horizon system of a quintom-dominated flat Universe, and \cite{GSL Interacting holographic DE} for various  interacting dark energy models.

Moreover,  the  GSL has  also been used as a validity constraint  on modified and alternative theories of gravity. For instance, the GSL has been imposed on the  event-horizon system of  the flat  Universe of $f(R)$ gravity in \cite{EventH GSL fR flat},
tentatively to the flat  apparent-horizon system of  generic modified gravities  in \cite{2nd law modified gravity},
to  the  higher-dimensional Gauss-Bonnet and Lovelock gravities in \cite{GSL MG Akbar},
to 
the Gauss-Bonnet, Randall-Sundrum and Dvali-Gabadadze-Porrati braneworlds in \cite{GSL Gauss-Bonnet braneworld},
the Horava-Lifshitz gravity in \cite{GSL Horava-Lifshitz},
$F(R,\mathcal{G})$ generalized Gauss-Bonnet gravity in \cite{EventH GSLT Gauss Bonnet}, $f(T)$ generalized teleparallel gravity in \cite{GSL f(T) gravity},
scalar-tensor-chameleon gravity in \cite{GSL scalar tensor chameleon}, and the self-interacting $f(R)$  gravity in \cite{GSLT f R interacting}.
Note that in the studies of GSLs,
the debatable ``local equilibrium assumption'' has been widely adopted which supposes that the matter content and the causal boundary in use (mainly the apparent or the event horizon) would have the same temperature  \cite{GSL Interacting holographic DE, GSL MG Akbar, GSL Gauss-Bonnet braneworld, GSL Horava-Lifshitz, GSL scalar tensor chameleon, EventH GSLT Gauss Bonnet, GSLT f R interacting}.

Unlike \emph{laboratory} thermodynamics which is a well-developed self-consistent framework, the  thermodynamics of the Universe is  practically a mixture of ordinary thermodynamics with analogous gravitational quantities, for which the consistency between the first and second laws and among the setups of thermodynamic functions are not yet verified. For example, the Hayward temperature  $\kappaup/2\pi$  \cite{Negative T=kappa/2pi I, Inverse brane world} or  $|\kappaup|/2\pi$  \cite{Cai IV, Inverse brane world T absolute} which formally resembles  the Hawking temperature of (quasi)stationary black holes \cite{Bekenstein-Hawking entropy} has been adopted in the first laws, while in GSLs both  $|\kappaup|/2\pi$ \cite{2nd law modified gravity, GSL MG Akbar, GSL Gauss-Bonnet braneworld, GSL scalar tensor chameleon, GSLT f R interacting} and  the Cai--Kim  temperature \cite{GSL Interacting holographic DE, GSL Horava-Lifshitz, EventH GSLT Gauss Bonnet} are used. Moreover, in existent literature we have noticed six questions regarding the gravitational thermodynamics of the Universe:
\begin{enumerate}
  \item[(1)] For the Cai--Kim and the Hayward temperatures, which one is more appropriate for the cosmological boundaries? By solving this \emph{temperature confusion}, the equations of total energy differential at the horizons could also be determined;
  \item[(2)] For the Bekenstein--Hawking entropy in GR and the Wald entropy in modified gravities, are they unconditionally applicable to both the cosmological apparent and the event horizons?
  \item[(3)] Is the standard second law  for the physical matter really ill-behaved and thus needs to be saved by the GSL? This constitutes the cosmological \emph{entropy confusion};
  \item[(4)] Is the artificial local equilibrium assumption really necessary for the GSL?
  \item[(5)] The region enveloped by the apparent horizon is actually a thermodynamically open system with the absolute cosmic Hubble flow crossing the horizon; how will this fact influence the entropy evolution?
  \item[(6)] Are the thermodynamic quantities fully consistent with each other when the cosmic gravitational thermodynamics is systemized?
\end{enumerate}
 In this paper, we will try to answer these questions.

This paper is organized as follows. Starting with GR and the $\Lambda$CDM Universe (where $\Lambda$ denotes generic dark energy), in Sec.~\ref{section FRW setups} we derive the holographic-style dynamical equations governing  the  apparent-horizon dynamics and the cosmic spatial expansion, which yield the  constraints from  the EoS parameter $w_m$ on the evolution and metric signature of the  apparent horizon. Section ~\ref{GR first laws Thermodynamics} demonstrates how these holographic-style gravitational equations imply the unified first law of thermodynamics and the Clausius equation, and shows the latter encodes the positive-heat-out sign convention for the horizon
entropy.  In Sec.~\ref{Solution to the horizon-temperature confusion} the Cai--Kim temperature is extensively compared with Hayward, with  the former chosen for further usage in Sec.~\ref{The (generalized) second laws of thermodynamics}, where we adjust the traditional Gibbs equation into the Positive Out convention to investigate the entropy evolution for the simple open systems enveloped by the apparent and event horizons. Finally  the whole framework of gravitational thermodynamics is extended from $\Lambda$CDM model and GR to generic modified gravity in
Sec.~\ref{Gravitational thermodynamics in ordinary modified gravities}.
Throughout this paper, we adopt the sign convention $\Gamma^\alpha_{\beta\gamma}=\Gamma^\alpha_{\;\;\,\beta\gamma}$, $R^{\alpha}_{\;\;\beta\gamma\delta}=\partial_\gamma \Gamma^\alpha_{\delta\beta}-\partial_\delta \Gamma^\alpha_{\gamma\beta}\cdots$ and  $R_{\mu\nu}=R^\alpha_{\;\;\mu\alpha\nu}$ with the metric signature $(-,+++)$.

\section{Dynamics of the cosmological apparent horizon}\label{section FRW setups}

\subsection{Apparent horizon and observable Universe}
The  FRW metric  provides the most general description for the spatially homogeneous and isotropic Universe. In the $(t,r,\theta,\varphi)$ coordinates for an observer comoving with the cosmic Hubble flow, it has the line element (e.g. \cite{Cai I, Bak Cosmic holography})
\begin{equation}\label{FRW metric I}
\begin{split}
ds^2 &= -dt^2+\frac{a(t)^2}{1-kr^2} dr^2 + a(t)^2 r^2 \big( d\theta^2+\sin^2 \!\theta d\varphi^2 \big)\\
&= h_{\alpha\beta} dx^\alpha dx^\beta+\Upsilon^2  \big( d\theta^2 +\sin^2 \!\theta  d\varphi^2 \big) ,
\end{split}
\end{equation}
where $a(t)$ refers to the scale factor to be specified by the gravitational field equations,
and the index $k$ denotes the normalized spatial curvature, with $k=\{+1\,,0\,,-1\}$  corresponding to closed, flat and open Universes, respectively.
$h_{\alpha\beta}\coloneqq\text{diag}[-1\,, \frac{a(t)^2}{1-kr^2}]$ represents the transverse two-metric spanned by  $x^\alpha=(t,r)$, and $\Upsilon\coloneqq a(t)\,r$ stands for the astronomical  circumference/areal radius.
Based on Eq.(\ref{FRW metric I}), one can establish the following null tetrad  adapted to the spherical symmetry and the null radial flow,
\begin{equation}\label{Tetrad I}
\begin{split}
\ell^\mu\,&=\,\bigg(\,1\,,\frac{\sqrt{1-kr^2}}{a}\,,0\,,0 \bigg)\\
n^\mu\,&=\,\frac{1}{2}\,\bigg(-1\,,\frac{\sqrt{1-kr^2}}{a}\,,0\,,0 \bigg)\\
m^\mu\,&=\,\frac{1}{\sqrt{2}\,\Upsilon}\,\bigg(0,0,1,\frac{i}{\sin\!\theta}\bigg)\,,
\end{split}
\end{equation}
which has been  adjusted to be compatible with the metric signature $(-,+++)$ (e.g.  Appendix B in \cite{Isolated Horizons Hamiltonian evolution}).
By calculating the Newman-Penrose spin coefficients $\rho_{\text{NP}} \coloneqq -m^\mu \bar{m}^\nu \nabla_\nu \ell_\mu$ and $\mu_{\text{NP}} \coloneqq \bar{m}^\mu m^\nu\nabla_\nu n_\mu$, the outward expansion rate $\theta_{(\ell)}= -\big(\rho_{\text{NP}}+\bar{\rho}_{\text{NP}}\big)$ and the inward expansion $\theta_{(n)}=\mu_{\text{NP}}+\bar{\mu}_{\text{NP}}$ are respectively found to be
\begin{equation}\label{expansion rate}
\begin{split}
\theta_{(\ell)}\,&=\,2H+2\Upsilon^{-1} \sqrt{1-\frac{k\Upsilon^2}{a^2}}\\
\theta_{(n)}\,&=\,-H+\Upsilon^{-1} \sqrt{1-\frac{k\Upsilon^2}{a^2}}\,,
\end{split}
\end{equation}
where $H$ refers to the time-dependent Hubble parameter of cosmic spatial expansion, and  $\displaystyle H\coloneqq\frac{\dot{a}}{a}$ with the overdot denoting the derivative with respect to the comoving time $t$.
For the expanding ($H>0$) Universe, $\theta_{(\ell)}$ and $\theta_{(n)}$
locate the apparent horizon $\Upsilon=\Upsilon_{\text{A}}$ by the unique
 marginally inner trapped horizon \cite{Trapping horizons} at
\begin{equation}\label{Horizon location}
\Upsilon_{\text{A}} \,=\,\frac{1}{\sqrt{H^2+\displaystyle \frac{k}{a^2}}}\,,
\end{equation}
with $\theta_{(\ell)}=4H>0$, $\theta_{(n)}=0$, and also $\partial_\mu \Upsilon$ becomes a null vector with $g^{\mu\nu}\partial_\mu \Upsilon\partial_\nu \Upsilon=0$ at  $\Upsilon_{\text{A}}$.
Immediately the temporal derivative of  Eq.(\ref{Horizon location}) yields the kinematic equation
\begin{equation}\label{dot Upsilon}
\dot{\Upsilon}_{\text{A}}\,=\,-H \Upsilon_{\text{A}}^3 \,\left( \dot{H}-\frac{k}{a^2}\right)\,.
\end{equation}
Just like $\Upsilon_{\text{A}}$ and $\dot{\Upsilon}_{\text{A}}$, hereafter quantities evaluated on or related to the apparent horizon will be highlighted by the subscript \textit{A}.

\{$\ell^\mu$, $n^\mu$\} in Eq.(\ref{Tetrad I}) coincide with the outgoing and ingoing tangent vector fields of the null radial congruence that is sent towards infinity by the comoving observer at $r=0$, and ingoing signals from the antitrapped region $\Upsilon>\Upsilon_{\text{A}}$ (where $\theta_{(\ell)}>0$, $\theta_{(n)}>0$) can no longer cross the marginally inner trapped $\Upsilon_{\text{A}}$ and return to the observer. However, the region  $\Upsilon\leq\Upsilon_{\text{A}}$ is not necessarily the standard  \emph{observable Universe} in astronomy where  ultrahigh redshift and visually superluminal recession can be detected \cite{common misconceptions, FRW studies}: $\Upsilon_{\text{A}}$ is a future-pointed horizon determined in active measurement by the observer, while the observable Universe is the past-pointed region measured by passive reception of distant signals and thus more related to the past particle horizon.

Note that we are working with the generic FRW metric Eq.(\ref{FRW metric I}) which allows for a nontrivial spatial curvature. This is not just for theoretical generality: in fact, astronomical observations indicate that the Universe may not be perfectly flat.
For example, in the $o\Lambda$CDM sub-model with a strict vacuum-energy condition $w_\Lambda=-1$,  the nine-years data from the Wilkinson Microwave Anisotropy Probe (WMAP) and other sources like the Baryon Acoustic Oscillations (BAO) yield the fractional energy density  $\Omega_k=-0.0027^{+0.0039}_{-0.0038}$ \cite{Nonflat WMAP 9} for the spatial curvature, independently the time-delay measurements of two strong gravitational lensing systems along with the seven-years
WMAP data find $\Omega_k=0.003^{+0.005}_{-0.006}$ \cite{Nonflat lensing}, while most recently analyses based on BAO data give $\Omega_k=-0.003\pm 0.003$ \cite{Nonflat BAO}.


\subsection{Holographic-style dynamical equations}\label{Dynamical equations for the apparent-horizon radius}


The matter content of the  Universe is usually portrayed by a  perfect-fluid type stress-energy-momentum tensor, and  in the metric-independent form it reads
\begin{equation}
\begin{split}
&T^{\mu\,(m)}_{\;\;\nu}=\text{diag}\big[-\rho_m,P_m,P_m,P_m\big]\\
&\text{with}\quad P_m/\rho_m\eqqcolon w_m \,,
\end{split}
\end{equation}
where $w_m$ refers to the equation of state (EoS) parameter.
Substituting this $T^{(m)}_{\mu\nu}$ and the metric Eq.(\ref{FRW metric I}) into Einstein's equation $R_{\mu\nu}-\frac{1}{2}Rg_{\mu\nu} =8\pi G T_{\mu\nu}^{(m)}$,  one obtains the  first and the second Friedmann equations
\begin{equation}\label{GR Friedmann eqn 1st}
\begin{split}
&H^2+\frac{k}{a^2}= \frac{8\pi G}{3}\rho_m  \quad\text{and}\\
\dot H-\frac{k}{a^2}&=-4\pi G\,\big(1+w_m\big)\,\rho_m=-4\pi Gh_m \\
\text{or}\quad &2\dot H+3H^2+\frac{k}{a^2}= -8\pi G P_m \,,
\end{split}
\end{equation}
where $h_m=\rho_m+P_m=\big(1+w_m\big)\rho_m$ refers to the enthalpy density.

Primarily, the first and second Friedmann equations are respectively the first and second order differential equations of the scale factor $a(t)$, which is the only unspecified function in the metric Eq.(\ref{FRW metric I}). On the other hand, recall the location and the time-derivative of the cosmological apparent horizon in Eqs. (\ref{Horizon location}) and (\ref{dot Upsilon}), and thus  Eq.(\ref{GR Friedmann eqn 1st})  can be rewritten into
\begin{equation}\label{GR Friedmann Upsilon}
\Upsilon_{\text{A}}^{-2}\,=\, \frac{8\pi G}{3}\,\rho_m  \,,
\end{equation}
\begin{equation}\label{GR Friedmann Upsilon II}
\dot{\Upsilon}_{\text{A}}
\,=\,4\pi G H \Upsilon_{\text{A}}^3 \big(1 + w_m \big)\, \rho_m
\,=\,4\pi GH \Upsilon_{\text{A}}^3 h_m\,,
\end{equation}
which manifest themselves as the dynamical equations of the apparent horizon. However, they also  describe the dynamics of spatial expansion for the entire Universe, so for this usage we will  dub Eqs.(\ref{GR Friedmann Upsilon}) and (\ref{GR Friedmann Upsilon II}) the ``holographic-style'' dynamical equations since they  reflect   the spirit of holography [we are using the word ``holographic'' in a generic sense  as opposed to the standard terminology \emph{holographic principle} in quantum gravity and string theory \cite{Holography I} or the holographic gravity method \cite{Padmanabhan holographic gravity}].

Eq.(\ref{GR Friedmann Upsilon}) immediately implies that, for the late-time Universe dominated by dark energy $\rho_m=\rho_\Lambda$,  the apparent horizon serves as  the natural infrared cutoff for the holographic dark energy model \cite{holographic dark energy Li}, in which the dark-energy density $\rho_\Lambda^{\text{(HG)}}$  relies on the scale of the infrared cutoff $\Upsilon_{\text{IR}}$  by $\rho_\Lambda^{\text{(HG)}}=3\Upsilon_{\text{IR}}^{-2}/(8\pi G)$.

Moreover, with the apparent-horizon area $A_{\text{A}}=4\pi \Upsilon_{\text{A}}^2$, it follows from  Eq.(\ref{GR Friedmann Upsilon}) that
\begin{equation}\label{GR Arho}
\rho_m A_{\text{A}} \,=\, \frac{3}{2 G}\,,
\end{equation}
so Eq.(\ref{GR Friedmann Upsilon II})  can be further simplified into
\begin{equation}\label{GR Friedmann Upsilon III}
\dot{\Upsilon}_{\text{A}}
\,=\,\frac{3}{2} H \Upsilon_{\text{A}} \,\big(1 + w_m \big)\,.
\end{equation}
With the help of Eqs.(\ref{GR Friedmann Upsilon}) and (\ref{GR Friedmann Upsilon III}),
for completeness  the third member (the  $P_m$ one) in Eq.(\ref{GR Friedmann eqn 1st}) can be directly translated into
\begin{equation}\label{GR Friedmann Upsilon IV}
\Upsilon_{\text{A}}^{-3}\,\Big( \dot{\Upsilon}_{\text{A}} -\frac{3}{2}H \Upsilon_{\text{A}}  \Big)\,=\,4\pi GHP_m\,,
\end{equation}
and we keep it in this form without further manipulations for later use in Sec.~\ref{GR Unified first law of thermodynamics}.

From a mathematical point of view, it might seem trivial to rewrite the Friedmann  equations (\ref{GR Friedmann eqn 1st}) into the holographic-style gravitational equations (\ref{GR Friedmann Upsilon})-(\ref{GR Friedmann Upsilon IV}).  However, considering that
existent studies on the gravitational thermodynamics of the cosmological apparent horizon always start from the relevant Friedmann equations  \cite{Cai V, Negative T=kappa/2pi I, Cai IV, Inverse brane world, Inverse brane world T absolute, scalar f R phi phi2, UFL Horava-Lifshitz,
causal boundary entropy, GSLT quintom, GSL Interacting holographic DE, EventH GSL fR flat,  2nd law modified gravity, GSL MG Akbar, GSL Gauss-Bonnet braneworld, GSL Horava-Lifshitz, EventH GSLT Gauss Bonnet, GSL f(T) gravity, GSL scalar tensor chameleon, GSLT f R interacting}, we wish that the manipulations of Eq.(7) into Eqs.(8)-(12) could make the formulations physically more meaningful and more concentrative on the horizon  $\Upsilon_{\text{A}}$ itself. Also, we will proceed to investigate some useful properties of the apparent horizon as necessary preparations for the horizon thermodynamics.

Eq.(\ref{GR Friedmann Upsilon III}) clearly shows that,  for an expanding Universe ($H>0$)  the apparent-horizon radius $\Upsilon_{\text{A}}$ can be either expanding, contracting or even static, depending on the domain of the EoS parameter $w_m$ or equivalently the sign of the enthalpy density $h_m$. In the $\Lambda$CDM cosmology, $\rho_m$ could be decomposed into all possible components,
$\rho_m=\sum \rho_{m}^{(i)}=\rho_m(\text{baryon})+\rho_m(\text{radiation})+\rho_m(\text{neutrino})
+\rho_m(\text{dark matter})+\rho_m(\text{dark energy})+\cdots,$
and the same for $P_m$. In principle there should be an EoS parameter $w_{m}^{(i)}
=P_{m}^{(i)}/\rho_{m}^{(i)}$ associated to each energy component. However, practically we can regard $w_m$ either as that of the absolutely dominating matter, or the weighted average for all relatively dominating components
\begin{equation}
w_m \,=\,  \frac{\sum P_{m}^{(i)}}{\rho_m} \,=\,  \frac{\sum w_{m}^{(i)} \rho_{m}^{(i)} }{\rho_m}
 \,=\, \sum \alpha_i\, w_{m}^{(i)}\,,
\end{equation}
 with the weight coefficient given by $\alpha_i=\rho_{m}^{(i)}/\rho_m$,
and thus $w_m$ varies over cosmic time scale. Then it follows from Eq.(\ref{GR Friedmann Upsilon III}) that:

\begin{widetext}
\begin{center}
 \renewcommand\arraystretch{1.26}
\newcommand{\tabincell}[2]{\begin{tabular}{@{}#1@{}}#2\end{tabular}}
\begin{tabular}{|c|c|c|l|}
  \hline
  $w_m$ & dominating matter  & enthalpy density & $\qquad\quad\dot{\Upsilon}_{\text{A}}$ \\   \hline\hline
  $\tabincell{c}{ $-1/3\leq w_m\,(\leq 1)$ and\\ $-1< w_m < -1/3$}$ & \tabincell{c}{ ordinary matter, and\\ quintessence \cite{DE Quintessence}} &$h_m>0$ & $\dot{\Upsilon}_{\text{A}}>0$, expanding \\ \hline
  $w_m=-1$  & \tabincell{c}{ cosmological constant or \\vacuum energy \cite{Dark Energy Reviews}}   &$h_m=0$ & $\dot{\Upsilon}_{\text{A}}=0$, static \\   \hline
  $w_m<-1$  & phantom \cite{DE phantom}  &$h_m<0$ & $\dot{\Upsilon}_{\text{A}}<0$, contracting  \\ \hline
\end{tabular}
\end{center}
\end{widetext}

The dominant energy condition \cite{Hawking Ellis} $\rho_m\geq |P_m|$
imposes the constraint $-1\leq w_m\leq 1$ for \emph{nonexotic} matter. Here we retain the upper limit $w_m\leq 1$ but loosen the lower limit, allowing $w_m$ to cross the barrier $w_m=-1$ into the \emph{exotic} phantom domain $w_m<-1$. The upper limit however is bracketed as $(\leq 1)$ to indicate that it is a physical rather than mathematical constraint.


\subsection{Induced metric of the apparent horizon}\label{GR Induced metric of the apparent horizon}
The total derivative of  $\Upsilon=\Upsilon(t,r)$ yields $adr=d\Upsilon-H\Upsilon dt$,
which recasts the FRW line element Eq.(\ref{FRW metric I}) into the $(t,\Upsilon,\theta,\varphi)$ coordinates as
\begin{eqnarray}
ds^2=\left(1-\frac{k\Upsilon^2}{a^2}\right)^{-1}&&\Bigg(
-\Big(1-\frac{\Upsilon^2}{\Upsilon_{\text{A}}^{2}}\Big)dt^2
- 2H\Upsilon  dtd\Upsilon
+ d\Upsilon^2 \Bigg)\nonumber\\
+\Upsilon^2 &&\Big( d\theta^2 +\sin^2 \!\theta  d\varphi^2 \Big).\label{FRW metric II}
\end{eqnarray}
Although  the comoving transverse coordinates $(t,r)$ are easier to work with, we will switch to the more physical coordinates $(t,\Upsilon)$ whenever necessary.
The metric  Eq.(\ref{FRW metric II}) reduces to become a three-dimensional hypersurface in the $(t,\theta,\varphi)$ coordinates at the apparent horizon $\Upsilon_{\text{A}}=\Upsilon_{\text{A}}(t)$,  and with Eq.(\ref{GR Friedmann Upsilon III}), the induced horizon metric turns out to be
\begin{eqnarray}
ds^2&=&\big(H\Upsilon_{\text{A}}\big)^{-2} \big(
\dot{\Upsilon}_{\text{A}} - 2H\Upsilon_{\text{A}}
\big)\dot{\Upsilon}_{\text{A}}  dt^2+\Upsilon_{\text{A}}^2 \big( d\theta^2 +\sin^2 \!\theta d\varphi^2 \big) \nonumber\\
&=&\frac{9}{4} \big( w_m  + 1\big) \big(
w_m  - \frac{1}{3}
\big)  dt^2+\Upsilon_{\text{A}}^2 \big( d\theta^2 +\sin^2 \!\theta  d\varphi^2 \big) .\label{FRW metric III}
\end{eqnarray}
Here $w_m$ shows up in the coefficients of $dt^2$, and indeed the spirit of geometrodynamics  allows and encourages physical parameters to directly participate in the spacetime metric, just like the mass, electric charge and angular momentum parameters in the Kerr-Newmann solution. It is easily seen that the signature of the apparent horizon solely relies on the domain of  $w_m$ regardless of the Universe being expanding or contracting.
\begin{enumerate}
  \item[(1)] For $-1<w_m<1/3$, the apparent horizon $\Upsilon_{\text{A}}$ has the signature $(-,++)$ and is timelike, which shares the signature of a quasilocal timelike membrane  in  black-hole physics \cite{Trapping horizons, Ivan Black hole boundaries}.
   \item[(2)] For $w_m<-1$ or $1/3<w_m\,(\leq 1)$,  the signature is $(+,++)$ and thus  $\Upsilon_{\text{A}}$  is spacelike. This situation has the same signature with the dynamical  black-hole horizons \cite{Dynamical horizon}.
  \item[(3)] For $w_m=-1$ or $w_m=1/3$,   $\Upsilon_{\text{A}}$ is a null surface with the signature $(0,++)$, so it coincides with the cosmological event horizon  $\Upsilon_{\text{E}}\coloneqq a\int_t^\infty a^{-1}d\hat{t}$ \cite{Bak Cosmic holography, Faraoni} which by definition is a future-pointed null causal boundary, and it shares the signature of isolated  black-hole horizons \cite{Isolated Horizons Hamiltonian evolution}.
\end{enumerate}
Note that these analogies between  $\Upsilon_{\text{A}}$ and black-hole horizons are limited to the metric signature, while the behaviors of their expansions \{$\theta_{(\ell)}$,  $\theta_{(n)}$\} and the horizon trappedness are entirely different.
Among the two critical  values,  $w_m=-1$ corresponds to the   de Sitter Universe dominated by a positive cosmological constant (or vacuum energy) \cite{de Sitter early 1}, while $w_m=1/3$ refers to the highly relativistic limit of $w_m$ and  the EoS of radiation, with  the trace of the the stress-energy-momentum tensor $g^{\mu\nu}T_{\mu\nu}^{(m)}=(3w_m-1)\rho_m$ vanishing at $w_m=1/3$. As will be shown later in Sec.~\ref{Solution to the horizon-temperature confusion},  $w_m=1/3$ also serves as the ``zero temperature divide'' if the apparent-horizon temperature were measured by $\kappaup/2\pi$ in terms of the Hayward surface gravity $\kappaup$.



\subsection{Relative evolution equations}

The  nontrivial $t$-component of
 $\nabla_\mu T^{\mu\,(m)}_{\;\;\nu}=0$  with respect to the  metric Eq.(\ref{FRW metric I}) leads to the continuity equation for the cosmic perfect fluid
\begin{equation}\label{GR continuity eqn}
 \dot{\rho}_m + 3 H\big(1+w_m  \big)\,\rho_m   \,=\,0\,.
\end{equation}
Thus for the relative evolution rate of the energy density $\dot{\rho}_m/\rho_m$, its
ratio over that of the cosmic scale factor $\dot{a}/a=H$ synchronizes with the instantaneous value of the EoS parameter, $\frac{\dot{\rho}_m}{\rho_m} \left/\frac{\dot{a}}{a}\right. = -3 \big(1+w_m  \big)$. This relation is not alone, as one could easily observe from Eq.(\ref{GR Friedmann Upsilon III}) that the relative evolution rate of the apparent-horizon radius $\dot{\Upsilon}_{\text{A}}/\Upsilon_{\text{A}}$
 is normalized by $\dot{a}/a$ into $\frac{\dot{\Upsilon}_{\text{A}}}{\Upsilon_{\text{A}}}\left/\frac{\dot{a}}{a}\right.=\frac{3}{2}(1 + w_m)$. These two equations reveal the interesting result that throughout the history of the Universe,  the relative evolution rate of the energy density is always proportional to  that of the apparent-horizon radius:
\begin{equation}\label{GR synchronization 3}
\frac{\dot{\rho}_m}{\rho_m}\,\bigg/\,\frac{\dot{\Upsilon}_{\text{A}}}{\Upsilon_{\text{A}}}   \,=\, -2\,.
\end{equation}
In fact, integration of Eq.(\ref{GR synchronization 3}) yields $\ln \rho_m\propto -2\ln \Upsilon_{\text{A}}$ and thus $\rho_m\propto \Upsilon_{\text{A}}^{-2}$, which matches the holographic-style dynamical equation (\ref{GR Friedmann Upsilon}) with the proportionality constant identified as $ \frac{3}{8\pi G}$.


\section{Thermodynamic implications of the holographic-style dynamical equations}\label{GR first laws Thermodynamics}

In Sec.~\ref{section FRW setups}, based on Eqs.(\ref{GR Friedmann Upsilon})-(\ref{GR Friedmann Upsilon IV}) we have analyzed some properties of the cosmological apparent horizon $\Upsilon_{\text{A}}$ to facilitate the subsequent discussion; one can refer to \cite{Faraoni} for more discussion of the horizon $\Upsilon_{\text{A}}$. From this section on, we will continue to investigate the thermodynamic implications of the holographic-style gravitational equations (\ref{GR Friedmann Upsilon})-(\ref{GR Friedmann Upsilon IV}).


\subsection{Unified first law of thermodynamics}\label{GR Unified first law of thermodynamics}

The  mass $M=\rho_mV$ of cosmic fluid within a sphere of radius $\Upsilon$, surface area
$A=4\pi \Upsilon^2$  and volume $V=\frac{4}{3}\pi \Upsilon^3$,  can be geometrically recovered  from the spacetime metric and we will identify it as the total internal energy $E$. With the Misner-Sharp mass/energy \cite{Misner-Sharp mass}  $E_{\text{MS}}\coloneqq \frac{\Upsilon }{2G}
 \big( 1- h^{\alpha\beta}\partial_\alpha \Upsilon\partial_\beta \Upsilon\big)$ for spherically symmetric spacetimes,  Eq.(\ref{FRW metric I}) with $h^{\alpha\beta}$$=\text{diag}[-1\,, \frac{a^2}{1-kr^2}]$ for the Universe yields
\begin{equation}\label{mass}
E\,=\,\frac{\Upsilon^3}{2G\Upsilon_{\text{A}}^2}\,,
\end{equation}
and its equivalence with the physically defined mass  $E=M=\rho_mV$ is guaranteed by Eq.(\ref{GR Friedmann Upsilon}). Equation (\ref{mass}) can also be reconstructed in the tetrad Eq.(\ref{Tetrad I}) from
the  Hawking energy \cite{Hawking mass} $E_{\text{Hk}}\coloneqq\frac{1}{4\pi G} \left( \int  \frac{dA}{4\pi} \right)^{1/2}\int \big( -\Psi_2-\sigma_{\text{NP}} \lambda_{\text{NP}} +\Phi_{11}+\Lambda_{\text{NP}} \big)dA
\equiv \frac{1}{4\pi G} \left( \int  \frac{dA}{4\pi} \right)^{1/2} \left( 2\pi-\int  \rho_{\text{NP}} \mu_{\text{NP}}dA \right)$ for twist-free spacetimes.
Immediately, the total derivative or transverse gradient of $E=E(t,r)$  is
\begin{eqnarray}
dE\,&=&-\frac{1}{G}\frac{ \Upsilon^3}{\Upsilon_{\text{A}}^3}\,\Big( \dot{\Upsilon}_{\text{A}}-\frac{3}{2}H \Upsilon_{\text{A}}  \Big)\,dt
+\frac{3}{2G} \frac{\Upsilon^2}{\Upsilon_{\text{A}}^2}  \,adr\label{GR dM inside t r u}\\
&=&-\frac{\dot{\Upsilon}_{\text{A}}}{G}\frac{\Upsilon^3}{\Upsilon_{\text{A}}^3} \,dt
+\frac{3}{2G} \frac{\Upsilon^2}{\Upsilon_{\text{A}}^2} \,d\Upsilon \label{GR dM inside t Upsilon u}\,,
\end{eqnarray}
where the relation $adr=d\Upsilon-H\Upsilon dt$ has been employed
to rewrite Eq.(\ref{GR dM inside t r u}) into Eq.(\ref{GR  dM inside t Upsilon u}), with the transverse coordinates from $(t,r)$ to $(t\,,\Upsilon)$.
According to the holographic-style dynamical equations  (\ref{GR Friedmann Upsilon}),  (\ref{GR Friedmann Upsilon II}) and   (\ref{GR Friedmann Upsilon IV}), the energy differentials Eqs.(\ref{GR dM inside t r u})  and (\ref{GR  dM inside t Upsilon u}) can be rewritten into
\begin{eqnarray}
dE\,&=&\,-A \Upsilon  HP_m\,dt+A\,\rho_m\, adr\label{GR  dE inside t r wu}\\
&=&-A\,\big(1+w_m\big)\rho_m \,H\Upsilon\,dt
+ A\,\rho_m\,d\Upsilon \label{GR dE inside t Upsilon wu}\,.
\end{eqnarray}
Eqs.(\ref{GR dE inside t r wu}) and (\ref{GR dE inside t Upsilon wu}) can be formally compactified into
\begin{eqnarray}\label{GR Unified first law}
dE\,=\, A\bm\psi+W dV\,,
\end{eqnarray}
where $\bm\psi$ and $W$ are respectively the energy supply covector
\begin{eqnarray}
\bm\psi\;&&=-\frac{1}{2} \,\rho_m\big(1 +w_m \big)\,H\Upsilon\,dt
+ \frac{1}{2} \,\rho_m\big(1 +w_m \big)\,adr\label{GR psi flux density t r w}\\
&&=\;\;- \, \,\rho_m\big(1 +w_m \big)\,H\Upsilon\,dt
+ \frac{1}{2} \,\rho_m\big(1 +w_m \big)\,d\Upsilon\label{GR psi flux density t Upsilon w}\,,
\end{eqnarray}
and the work density
\begin{equation}\label{GR work density}
W
\,=\,\frac{1}{2}\big(1 -w_m \big) \,\rho_m\,.
\end{equation}
Eq.(\ref{GR Unified first law}) is exactly the unified first law of (equilibrium) thermodynamics  proposed by Hayward \cite{Hayward Unified first law}, and one can see from the derivation process that it  applies to a volume of  arbitrary areal radius $\Upsilon$, no matter $\Upsilon < \Upsilon_{\text{A}}$,  $\Upsilon= \Upsilon_{\text{A}}$ or  $\Upsilon > \Upsilon_{\text{A}}$. Moreover, $W$ and $\bm\psi$ can respectively be traced back to the  scalar invariant $W \coloneqq -\frac{1}{2} T^{\alpha\beta}_{(m)} h_{\alpha\beta}$ and the covector  invariant $\bm\psi_\alpha \coloneqq T_{\alpha\,(m)}^{\;\;\,\beta} \partial_\beta \Upsilon+W \partial_\alpha \Upsilon$ \cite{Hayward Unified first law},
which are valid for all spherically symmetric spacetimes besides FRW, and have Eqs.(\ref{GR psi flux density t r w}), (\ref{GR psi flux density t Upsilon w}) and (\ref{GR work density}) as their concrete components with respect to the metric Eq.(\ref{FRW metric I}).

Note that the ``unified'' first law Eq.(\ref{GR Unified first law}) for the gravitational thermodynamics of the Universe is totally different from the first laws in black-hole thermodynamics which balance the energy differential with the first-order variations  of the Arnowitt-Deser-Misner type quantities (such as mass, electric charge, and angular momentum). Instead, Eq.(\ref{GR Unified first law}) is more related to the geometrical aspects of the thermodynamics-gravity correspondence.


\subsection{Clausius equation on the apparent horizon for an isochoric process}\label{GR Clausius equation On the horizon}

Having seen that the full set of holographic-style dynamical equations  (\ref{GR Friedmann Upsilon}),  (\ref{GR Friedmann Upsilon II}) and   (\ref{GR Friedmann Upsilon IV})
yield the unified first law $dE= A\bm\psi+W dV$ for an arbitrary region in the FRW Universe,   we will focus on the volume enclosed by the  apparent horizon $\Upsilon_{\text{A}}$. Firstly, Eq.(\ref{GR Friedmann Upsilon II}) leads to
\begin{equation}
\frac{ \dot{\Upsilon}_{\text{A}}}{G} dt
\; {=}\;A_{\text{A}}\,\big(1+w_m\big)\,\rho_m\,H\Upsilon_{\text{A}}\,dt\,,
\end{equation}
and the left  hand side  can be manipulated into
\begin{equation}\label{GR On horizon step I}
\frac{ \dot{\Upsilon}_{\text{A}}}{G} dt\,=\,\frac{1}{2 \pi \Upsilon_{\text{A}}}\left(\frac{ 2 \pi \Upsilon_{\text{A}}\dot{\Upsilon}_{\text{A}}}{G } dt \right)\,=\,
\frac{1}{2 \pi \Upsilon_{\text{A}}}\frac{d}{dt}\left(\frac{\pi \Upsilon_{\text{A}}^2 }{G }\right)\,.
\end{equation}
Applying the geometrically defined Hawking-Bekenstein entropy \cite{Bekenstein-Hawking entropy}  (in the units $\hbar= c = k\,$[Boltzmann] = 1) to the apparent horizon
\begin{equation}\label{GR Hawking-Bekenstein entropy}
S_{\text{A}} \;{=}\;\frac{\pi \Upsilon_{\text{A}}^2 }{G }\;=\;\frac{A_{\text{A}}}{4G }  \,,
\end{equation}
then  employing the Cai--Kim temperature \cite{Cai I, Temperature tuneling}
\begin{equation}\label{T}
T_{\text{A}}\,\equiv \,\frac{1}{2 \pi \Upsilon_{\text{A}}}\,,
\end{equation}
thus $T_{\text{A}}dS_{\text{A}}=   \dot{\Upsilon}_{\text{A}}/G dt$ and Eq.(\ref{GR On horizon step I}) can be rewritten into
\begin{equation}\label{GR TdS nonequi}
T_{\text{A}}dS_{\text{A}}\; {=}\;\delta Q_{\text{A}} \;{=}\;-  A_{\text{A}}\bm\psi_t \;{=}\;-dE_{\text{A}}\Big|_{d\Upsilon=0}\,,
\end{equation}
where $\bm\psi_t$ is the $t$-component of the energy supply covector $\bm\psi=\bm\psi_t+\bm\psi_\Upsilon=\psi_\alpha dx^\alpha$ in Eq.(\ref{GR psi flux density t Upsilon w}). This basically reverses Cai and Kim's formulation in \cite{Cai I}, and differs from \cite{Cai V} by the setup of the horizon temperature. Eq.(\ref{GR TdS nonequi}) is actually the Clausius equation for equilibrium and reversible thermodynamic processes, and the meaning of reversibility compatible with the cosmic dynamics is clarified in Appendix~\ref{The minimum set of state functions  and reversibility}. Comparing  Eq.(\ref{GR TdS nonequi}) with the unified first law Eq.(\ref{GR Unified first law}), one could find that
Eq.(\ref{GR TdS nonequi}) is just Eq.(\ref{GR Unified first law})
with the two $d\Upsilon$ components removed and then  evaluated at   $\Upsilon_{\text{A}}$. Assuming that the apparent horizon locates at $\Upsilon_{\text{A}0}\equiv\Upsilon_{\text{A}}(t=t_0)$ at an arbitrary moment $t_0$, then during the infinitesimal time interval $dt$ the horizon will move to $\Upsilon_{\text{A}0} +\dot{\Upsilon}_{\text{A}0} dt$; meanwhile, for the isochoric process of the volume $V(\Upsilon_{\text{A}0})$ (i.e. a ``controlled volume''), the amount of energy across the horizon $\Upsilon_{\text{A}0} $ is just $dE_{\text{A}}=A_{\text{A}0}\bm\psi_t$ evaluated at $t_0$, and for brevity we will drop the subscript ``0'' whenever possible as $t_0$ is arbitrary.

The energy-balance equation (\ref{GR TdS nonequi}) implies that the region  $\Upsilon\leq\Upsilon_{\text{A}}$ enveloped by the cosmological apparent horizon is thermodynamically an \emph{open} system which exchanges both heat and matter (condensed components in the Hubble flow) with its surroundings/reservoir  $\Upsilon\geq\Upsilon_{\text{A}}$.
Here we emphasize again that $\Upsilon_{\text{A}}$  is simply a visual boundary preventing ingoing null radial signals from reaching the comoving observer,
and the absolute cosmic Hubble flow can still cross $\Upsilon_{\text{A}}$.
Also, unlike nonrelativistic thermodynamics in which $\delta Q$ exclusively refers to the heat transfer (i.e. electromagnetic flow), the $\delta Q_{\text{A}}$ in  Eq.(\ref{GR TdS nonequi}) is used in a mass-energy-equivalence sense and denotes the Hubble energy flow which generally contains different matter components.

Finally,  for the open  system enveloped by $\Upsilon_{\text{A}}$,  we  combine the Clausius equation (\ref{GR TdS nonequi}) and the unified first law Eq.(\ref{GR Unified first law}) into the total energy differential
\begin{equation}\label{GR total dE Cai}
\begin{split}
dE_{\text{A}}\;&{=}\;A_{\text{A}}\psi_t\, dt+A_{\text{A}}\big(\psi_\Upsilon +W\big)\,d\Upsilon_{\text{A}}\\
&{=}\;-T_{\text{A}}dS_{\text{A}}+\rho_m\,A_{\text{A}}d\Upsilon_{\text{A}}\\
&{=}\; -T_{\text{A}}dS_{\text{A}}+\rho_m\,dV_{\text{A}}\,.
\end{split}
\end{equation}
In fact, by the continuity equation (\ref{GR continuity eqn}) one can verify $-T_{\text{A}}dS_{\text{A}}=V_{\text{A}}d\rho_m $, which agrees with the thermodynamic connotation that the heat $-T_{\text{A}}d S_{\text{A}}=\delta Q_{\text{A}}$ measures the loss of internal energy that can no longer  be used to do work. In this sense, one may further regard $dE_{\text{A}}+T_{\text{A}}dS_{\text{A}}$ to play the role of the relativistic differential Helmholtz free energy $d\mathds{F}_{\text{A}}$ for the instantaneous $\Upsilon_{\text{A}0} $ of temperature  $T_{\text{A}0} $,
\begin{equation}\label{dF Cai--Kim}
\begin{split}
d\mathds{F}_{\text{A}}\;\coloneqq\; dE_{\text{A}}+T_{\text{A}}dS_{\text{A}}
\;{=}\;\rho_m\,dV_{\text{A}}
\;{=}\;\left(\psi_\Upsilon +W\right)\,dV_{\text{A}}\,,
\end{split}
\end{equation}
which represents the maximal work element that can be extracted from the interior of  $\Upsilon_{\text{A}0} $; one could also identify the relativistic differential Gibbs free energy $d\mathds{G}_{\text{A}}$, which means the ``useful'' work element, as
\begin{equation}\label{dG Cai--Kim}
\begin{split}
d\mathds{G}_{\text{A}}\,\coloneqq \, dE_{\text{A}}+T_{\text{A}}dS_{\text{A}}+P_m dV_{\text{A}}
=\rho_m\left(1+w_m \right)dV_{\text{A}} \,.
\end{split}
\end{equation}
Note that $d\mathds{F}_{\text{A}}$ and $d\mathds{G}_{\text{A}}$ contain $+T_{\text{A}}dS_{\text{A}}$ with a plus instead of a minus sign, because the Cai--Kim Clausius relation $dE_{\text{A}}=-T_{\text{A}}dS_{\text{A}}$ encodes that the horizon entropy $S_{\text{A}}$ is defined in a ``positive heat out'' rather than the traditional positive-heat-in thermodynamic sign convention, as will be extensively discussed in Sec.~\ref{subsection Positive heat out thermodynamic sign convention}.


\section{Solution to the horizon-temperature confusion}\label{Solution to the horizon-temperature confusion}

\subsection{The horizon-temperature confusion}\label{The horizon-temperature confusion}

In the thermodynamics of (quasi)stationary black holes \cite{Bekenstein-Hawking entropy}, the  Hawking  temperature satisfies $T=\tilde{\kappa}/(2\pi)$  based on  the traditional Killing  surface gravity $\tilde{\kappa}$ and the Killing generators of the horizon.
For the FRW Universe,  one has the Hayward inaffinity parameter $\kappaup$ \cite{Hayward Unified first law} in place of the Killing inaffinity, which yields the  Hayward  surface gravity on the apparent horizon,
\begin{equation}\label{Hayward surface gravity}
\begin{split}
\kappaup\,&\coloneqq\, \frac{1}{2}h^{\alpha\beta}\nabla_\alpha \nabla_\beta \Upsilon= \frac{1}{2\sqrt{-h}}\partial_\alpha\Big(\!\sqrt{-h}\,h^{\alpha\beta}\partial_\beta \Upsilon \Big)\\
&\equiv\,-\frac{\Upsilon}{\Upsilon_{\text{A}}^2}\left(1-\frac{\dot{\Upsilon}_{\text{A}}}{2H\Upsilon_{\text{A}}} \right)
=-\frac{1}{\Upsilon_{\text{A}}}\left(1-\frac{\dot{\Upsilon}_{\text{A}}}{2H\Upsilon_{\text{A}}} \right)\Big|_{\Upsilon_{\text{A}}},
\end{split}
\end{equation}
where $h_{\alpha\beta}=\text{diag}[-1\,, \frac{a^2}{1-kr^2}]$  refers to the transverse two-metric in Eq.(\ref{FRW metric I}). Then formally following the Hawking temperature, the Hayward  temperature of  the  apparent horizon $\Upsilon_{\text{A}}$ is defined either by  \cite{Negative T=kappa/2pi I, Inverse brane world}
\begin{equation}\label{T complicated}
\mathcal{T}_{\text{A}}\,\coloneqq\,\frac{\kappaup}{2\pi}\,=\,-\frac{1}{2\pi\Upsilon_{\text{A}}}\left(1-\frac{\dot{\Upsilon}_{\text{A}}}{2H\Upsilon_{\text{A}}} \right)
\end{equation}
or \cite{Cai IV, Inverse brane world T absolute, 2nd law modified gravity, GSL MG Akbar, GSL Gauss-Bonnet braneworld, GSL scalar tensor chameleon, GSLT f R interacting}
\begin{equation}\label{T complicated absolute}
\mathcal{T}_{\text{A}}^{(+)}\,\coloneqq\,\frac{(\kappaup\,|}{2\pi}\,=\,\frac{1}{2\pi\Upsilon_{\text{A}}}\left(1-\frac{\dot{\Upsilon}_{\text{A}}}{2H\Upsilon_{\text{A}}} \right)\,,
\end{equation}
where we use the symbol $(\kappaup\,|$ to denote  the \emph{partial} absolute value of $\kappaup$, because existing papers  have \emph{a priori} abandoned  the possibility of $\dot{\Upsilon}_{\text{A}}/(2H\Upsilon_{\text{A}})\geq 1$ for $\mathcal{T}_{\text{A}}^{(+)}$.
Equation (\ref{T complicated absolute}) is always supplemented by the assumption   \cite{Cai IV, Inverse brane world T absolute, 2nd law modified gravity, GSL MG Akbar, GSL Gauss-Bonnet braneworld, GSL scalar tensor chameleon, GSLT f R interacting}
\begin{equation}\label{T complicated supplement}
\frac{\dot{\Upsilon}_{\text{A}}}{2H\Upsilon_{\text{A}}}\,<\,1
\end{equation}
to guarantee a positive $\mathcal{T}_{\text{A}}^{(+)}$ which is required by the third law of thermodynamics, and even the condition  \cite{2nd law modified gravity}
\begin{equation}\label{T complicated supplement II}
\frac{\dot{\Upsilon}_{\text{A}}}{2H\Upsilon_{\text{A}}}\,<<\, 1
\end{equation}
so that $\mathcal{T}_{\text{A}}^{(+)}$ can be approximated into the Cai--Kim temperature \cite{Cai I, Temperature tuneling}
\begin{equation}\label{T complicated supplement III}
\mathcal{T}_{\text{A}}^{(+)}\,\approx\,\frac{1}{2\pi\Upsilon_{\text{A}}}\;=\,{T}_{\text{A}}\,.
\end{equation}

Historically  the inverse problem ``from thermodynamics to gravitational equations for the Universe''  \cite{Cai I} was formulated earlier, in which  the Cai--Kim temperature works perfectly for all theories of gravity. Later on,  the problem ``from FRW gravitational equations to thermodynamics'' \cite{Cai V, Negative T=kappa/2pi I, Cai IV} (as the logic in this paper) came into attention in which the Hayward temperature seems to become effective. Considering that two different temperatures make the two mutually inverse problems asymmetric,  attempts have been made to reduce the differences between them, mainly the assumptions Eqs.(\ref{T complicated supplement}) and (\ref{T complicated supplement II}).

Note that when the conditions Eqs.(\ref{T complicated supplement II}) and (\ref{T complicated supplement III}) are applied to Eq.(\ref{T complicated}), $\mathcal{T}_{\text{A}}$ would become a negative temperature.  \cite{Negative T=kappa/2pi I} has suggested that it might be possible to understand this phenomenon as a consequence of the cosmological apparent horizon being inner trapped [$\theta_{(\ell)}>0$, $\theta_{(n)}=0$], as opposed to the positive temperatures of black-hole apparent horizons which are always marginally outer trapped [$\theta_{(\ell)}=0$, $\theta_{(n)}<0$]. However, this proposal turns out to be inappropriate; as will be shown at the end of Sec.~\ref{subsection zero temperature divide}, the signs of $\mathcal{T}_{\text{A}}$ actually keep pace with the metric signatures rather than the inner/outer trappedness of the horizon  $\Upsilon_{\text{A}}$.

\subsection{Effects of $\mathcal{T}_{\text{A}}dS_{\text{A}}$ and $\mathcal{T}_{\text{A}}^{(+)} dS_{\text{A}}$}

In Sec.~\ref{GR Clausius equation On the horizon}, we have seen $T_{\text{A}}dS_{\text{A}}=A_{\text{A}} \bm\psi_t$ for the Cai--Kim  $T_{\text{A}}=1/(2\pi \Upsilon_{\text{A}})$, and now let's examine  the effects of $\mathcal{T}_{\text{A}}$ and  $\mathcal{T}_{\text{A}}^{(+)}$.
Given the Bekenstein--Hawking entropy $S_{\text{A}}=A_{\text{A}}/4G$, the dynamical equation $\dot{\Upsilon}_{\text{A}}=A_{\text{A}}H \Upsilon_{\text{A}}  G\big(1 +w_m \big)\rho_m$ and the energy supply covector  $\bm\psi=\bm\psi_t
+ \bm\psi_\Upsilon =-   \big(1 +w_m \big)\rho_m H\Upsilon_{\text{A}}dt
+ \frac{1}{2} \big(1 +w_m \big)\rho_m\,d\Upsilon_{\text{A}}$, one has
\begin{eqnarray}
\mathcal{T}_{\text{A}}dS_{\text{A}}
&=&-\frac{ \dot{\Upsilon}_{\text{A}}}{G }+\frac{\dot{\Upsilon}_{\text{A}}}{2GH\Upsilon_{\text{A}}}\dot{\Upsilon}_{\text{A}}dt \nonumber\\
&=&-A_{\text{A}} H \Upsilon_{\text{A}} \big(1 +w_m \big)\rho_m dt
+\frac{1}{2} A_{\text{A}}   \big(1 +w_m \big)\rho_m d\Upsilon_{\text{A}}\nonumber\\
&=&\;\;A_{\text{A}} \bm\psi_t
+ A_{\text{A}} \bm\psi_\Upsilon=A_{\text{A}} \bm\psi \label{TdS complicated} \,.
\end{eqnarray}
 Similarly,
for the $\mathcal{T}_{\text{A}}^{(+)}$ defined in Eq.(\ref{T complicated absolute}),
\begin{equation}\label{TdS complicated absolute}
\mathcal{T}_{\text{A}}^{(+)} dS_{\text{A}}\,=\,-\Big(A_{\text{A}} \bm\psi_t
+ A_{\text{A}} \bm\psi_\Upsilon\Big)\,=\,-A_{\text{A}} \bm\psi \,.
\end{equation}
Hence, for the two terms comprising $\mathcal{T}_{\text{A}}$ and $\mathcal{T}_{\text{A}}^{(+)}$, the $\pm \frac{1}{2\pi\Upsilon_{\text{A}}}dS_{\text{A}}$  is balanced by $\mp A_{\text{A}} \bm\psi_t$, while the $\pm \frac{\dot{\Upsilon}_{\text{A}}}{2H\Upsilon_{\text{A}}}dS_{\text{A}}$  is equal to $\pm A_{\text{A}} \bm\psi_\Upsilon$. As obtained in e.g. \cite{Cai V}-\cite{UFL Horava-Lifshitz}, for the open system enveloped by the cosmological apparent horizon, combining Eqs.(\ref{TdS complicated}) and (\ref{TdS complicated absolute})
with the unified first law Eq.(\ref{GR Unified first law})  leads to the total energy differential
\begin{equation}\label{dE HK TdS}
\begin{split}
dE_{\text{A}}\;&=\;\;\;\;\mathcal{T}_{\text{A}}dS_{\text{A}}+WdV_{\text{A}}\\
&=-\mathcal{T}_{\text{A}}^{(+)} dS_{\text{A}}+WdV_{\text{A}}\,,
\end{split}
\end{equation}
as opposed to $dE_{\text{A}}=-T_{\text{A}}dS_{\text{A}}+\rho_m dV_{\text{A}}$ for the Cai--Kim $T_{\text{A}}$.

\subsection{``Zero temperature divide'' $w_m=1/3$ and preference of Cai--Kim temperature}\label{subsection zero temperature divide}

Now apply the dynamical equation (\ref{GR Friedmann Upsilon III}) to  \{$\mathcal{T}_{\text{A}}$, $\mathcal{T}_{\text{A}}^{(+)}$\} and the assumptions  in Eqs.(\ref{T complicated supplement}) and (\ref{T complicated supplement II}). With $\dot{\Upsilon}_{\text{A}}
=\frac{3}{2} H \Upsilon_{\text{A}} \,\big(1 + w_m \big)$, the Hayward surface gravity becomes
\begin{equation}\label{kappaup in wm}
\begin{split}
\kappaup
\;=\;
-\frac{1}{\Upsilon_{\text{A}}}\left(1-\frac{\dot{\Upsilon}_{\text{A}}}{2H\Upsilon_{\text{A}}} \right)
\;=\;-\frac{3}{4\Upsilon_{\text{A}}}\big( \frac{1}{3}-w_m \big)\,,
\end{split}
\end{equation}
so it follows that
\begin{equation}
\left\{ \begin{aligned}
&w_m> \frac{1}{3}: \quad  \kappaup>0\,, \quad |\,\kappaup\,|=\frac{3}{4\Upsilon_{\text{A}}}\big(w_m -\frac{1}{3}  \big)\\
&w_m= \frac{1}{3}: \quad \kappaup=|\,\kappaup\,|=0\\
&w_m< \frac{1}{3}: \quad \kappaup<0 \,, \quad |\,\kappaup\,|=\frac{3}{4\Upsilon_{\text{A}}}\big(\frac{1}{3}-  w_m  \big)
                          \end{aligned} \right.\,.
\end{equation}
The  Hayward temperature   $\mathcal{T}_{\text{A}}$ in Eq.(\ref{T complicated})  and its partially absolute value $\mathcal{T}_{\text{A}}^{(+)}$   in Eq.(\ref{T complicated absolute}) become
\begin{equation}\label{T complicated w}
\begin{split}
\mathcal{T}_{\text{A}}\,&
=-\frac{3}{8\pi\Upsilon_{\text{A}}}\big(\frac{1}{3}-  w_m  \big)
=-\frac{1}{4}T_{\text{A}}\big(1-  3w_m  \big)\\
\mathcal{T}_{\text{A}}^{(+)}\,&
=\;\;\frac{3}{8\pi\Upsilon_{\text{A}}}\big(\frac{1}{3}-  w_m  \big)
=\;\;\frac{1}{4}T_{\text{A}}\big(1-  3 w_m  \big)\,.
\end{split}
\end{equation}
Fortunately $\mathcal{T}_{\text{A}}$ and  $\mathcal{T}_{\text{A}}^{(+)}$ remain as state functions, although Eqs.(\ref{T complicated}) and (\ref{T complicated absolute}) carry $\{\dot{\Upsilon}_{\text{A}}, H\}$ and look like process quantities (see Appendix~\ref{The minimum set of state functions  and reversibility} for more discussion). Moreover,
the supplementary assumption Eq.(\ref{T complicated supplement}) for $\mathcal{T}_{\text{A}}^{(+)}>0$ turns out to be
\begin{equation}\label{T complicated supplement w}
\frac{\dot{\Upsilon}_{\text{A}}}{2H\Upsilon_{\text{A}}}\,=\,\frac{3}{4} \big(1 + w_m \big) <1\quad\Rightarrow\quad
w_m<1/3\,.
\end{equation}
Thus the condition $\frac{\dot{\Upsilon}_{\text{A}}}{2H\Upsilon_{\text{A}}}\ll 1$ in Eq.(\ref{T complicated supplement II}) could be directly translated into $w_m\ll 1/3$, which is however inaccurate: in fact,  if directly starting from Eq.(\ref{T complicated w}),
the approximation  $\mathcal{T}_{\text{A}}^{(+)}\approx T_{\text{A}} = 1/(2\pi\Upsilon_{\text{A}})$ will require
\begin{equation}\label{T correct approx w}
w_m\,\to\,-1\,.
\end{equation}
It is neither mathematically nor physically identical with $w_m\ll 1/3$  which could only be  perfectly  satisfied for   $w_m\to-\infty$ in the extreme phantom domain.

Eqs.(\ref{kappaup in wm}) -- (\ref{T correct approx w}) have rewritten and simplified the original expressions of the Hayward  temperatures \{$\mathcal{T}_{\text{A}}$ , $\mathcal{T}_{\text{A}}^{(+)}$\} in Eqs.(\ref{T complicated}, \ref{T complicated absolute}) and their supplementary conditions Eqs.(\ref{T complicated supplement}, \ref{T complicated supplement II}). Based on these results we realize that it becomes possible to make  an extensive comparison between  \{$\mathcal{T}_{\text{A}}$ , $\mathcal{T}_{\text{A}}^{(+)}$\} and the Cai--Kim $T_{\text{A}}=1/(2\pi\Upsilon_{\text{A}})$, which reveals  the following facts.

\begin{enumerate}

  \item[(1)] $\mathcal{T}_{\text{A}}$ is negative definite  for $1/3<w_m\;(\leq 1)$, positive definite for $w_m<1/3$, and $\mathcal{T}_{\text{A}}\equiv 0$ for $w_m=1/3$. We will dub the special value $w_m=1/3$ as the  Hayward  ``zero temperature divide'', which is inspired by the terminology ``phantom divide'' for $w_m=-1$ in dark-energy physics \cite{Dark Energy Reviews}. Hence, $\mathcal{T}_{\text{A}}$ does not respect the third law of thermodynamics. Moreover, one has $\mathcal{T}_{\text{A}}=0$ at  $w_m= 1/3$ and thus
  $\mathcal{T}_{\text{A}}dS_{\text{A}}= 0$;
  following Eq.(\ref{TdS complicated}), this can be verified by
  \begin{eqnarray}
A_{\text{A}}\bm\psi&=&-A_{\text{A}} H \Upsilon_{\text{A}} \big(1 +w_m \big)\rho_m dt
+\frac{1}{2} A_{\text{A}}   \big(1 +w_m \big)\rho_m d\Upsilon_{\text{A}}\nonumber\\
&=&A_{\text{A}}\rho_m \big(1 +w_m \big)\big(\frac{1}{2}\dot{\Upsilon}_{\text{A}}-H\Upsilon_{\text{A}}\big) dt\nonumber\\
&=&\frac{9}{8G} H\Upsilon_{\text{A}} \big(1 +w_m \big)\big(w_m-\frac{1}{3}\big) dt .
      \end{eqnarray}

  \item[(2)] The condition $w_m<1/3$ for the validity of  $\mathcal{T}_{\text{A}}^{(+)}$ is too restrictive and unnatural, because $w_m=1/3$ serves as the EoS of radiation and $(1\geq)\,w_m> 1/3$ represents all  highly relativistic  energy components.  For example, it is well known that a canonical and homogeneous scalar field $\phi(t)$ in the FRW Universe has the EoS (e.g. \cite{Cai I})
   \begin{eqnarray}
w_m^{(\phi)}=\frac{P_\phi}{\rho_\phi}=\frac{\frac{1}{2}\dot{\phi}^2-V(\phi)}{\frac{1}{2}\dot{\phi}^2+V(\phi)}.
      \end{eqnarray}
$w_m^{(\phi)}$ can fall into the domain $1/3\leq w_m^{(\phi)}\leq1$ when the dynamical term $\frac{1}{2}\dot{\phi}^2$ dominates over the potential $V(\phi)$, and we donot see any physical reason to  \emph{a priori} rule out this kind of  fast-rolling scalar field.

  \item[(3)] The equality
$\mathcal{T}_{\text{A}}dS_{\text{A}}=A_{\text{A}}\big(\bm\psi_t
+ \bm\psi_\Upsilon \big)=-\mathcal{T}_{\text{A}}^{(+)} dS_{\text{A}}$ implies that  $\mathcal{T}_{\text{A}}dS_{\text{A}}$ and $\mathcal{T}_{\text{A}}^{(+)} dS_{\text{A}}$ need to be balanced by $dt$ and also the $d\Upsilon_{\text{A}}$ component of $\bm\psi$, and thus  the other $d\Upsilon_{\text{A}}$ component from $WdV_{\text{A}}=WA_{\text{A}}d\Upsilon_{\text{A}}$ should be nonvanishing as well. Hence, $\mathcal{T}_{\text{A}}dS_{\text{A}}$ and $\mathcal{T}_{\text{A}}^{(+)} dS_{\text{A}}$ always live together with $WdV_{\text{A}}$ to form the total energy differential Eq.(\ref{dE HK TdS}) rather than some Clausius-type equation $\delta \widetilde{Q}=\mathcal{T}_{\text{A}}^{(+)} dS_{\text{A}}=-\mathcal{T}_{\text{A}} dS_{\text{A}}=-A_{\text{A}} \bm\psi$, and there exists no isochoric process ($d\Upsilon=0$) for \{$\mathcal{T}_{\text{A}}$ , $\mathcal{T}_{\text{A}}^{(+)}$\}.

  \item[(4)]  The ``highly relativistic limit'' $w_m=1/3$ is more than the divide for negative, zero or positive Hayward temperature $\mathcal{T}_{\text{A}}$; it is also the exact divide for the induced metric of the apparent horizon to be spacelike, null or timelike, as discussed before in Sec.~\ref{GR Induced metric of the apparent horizon}. That is to say, the sign of the temperature synchronizes with the signature of the horizon metric. However, there are no such behaviors for analogies in black-hole physics: for example,
  a slowly-evolving  quasilocal black-hole  horizon \cite{Ivan Black hole boundaries, Slowly evolving horizons} can be either  spacelike, null or timelike, but the horizon temperature is always positive definite regardless of the horizon signature.

  \item[(5)] Unlike the Cai--Kim temperature  $T_{\text{A}}$, the Hayward \{$\mathcal{T}_{\text{A}}$, $\mathcal{T}_{\text{A}}^{(+)}$\} used for the problem ``from gravitational equations to thermodynamic relations for the Universe'' do not work for the problem ``from thermodynamic relations to gravitational equations''. That is to say, \{$\mathcal{T}_{\text{A}}$ , $\mathcal{T}_{\text{A}}^{(+)}$\} break the symmetry between the formulations of these two mutually inverse problems.


\end{enumerate}

On the other hand,  the Cai--Kim temperature $T_{\text{A}}=1/(2\pi \Upsilon_{\text{A}})$ is positive definite throughout the history of the Universe, it provides symmetric formulations of the conjugate problems ``gravity to thermodynamics'' and ``thermodynamics to gravity'', and it is the Hawking-like temperature measured by a Kodama observer for the matter tunneling into the untrapped interior $\Upsilon<\Upsilon_{\text{A}}$ from the antitrapped exterior $\Upsilon>\Upsilon_{\text{A}}$  \cite{Temperature tuneling}. In fact, besides the assumption Eq.(\ref{T complicated supplement II}) for the approximation $\mathcal{T}_{\text  A}^{(+)}\approx T_{\text{A}}$ in Eq.(\ref{T complicated supplement III}), there have been efforts to redefine  the  \emph{dynamical}  surface gravity in place of Eq.(\ref{Hayward surface gravity}) for the dynamical apparent horizon $\Upsilon_{\text{A}}$; for example, inspired by the thermodynamics of dynamical black-hole horizons \cite{Dynamical horizon}, the inaffinity $ \kappaup\coloneqq -\frac{1}{2}\partial_\Upsilon \Xi$ with $\Xi\coloneqq h^{\alpha\beta}\partial_\alpha \Upsilon\partial_\beta \Upsilon\equiv 1-\Upsilon^2/\Upsilon_{\text{A}}^2$ has been employed for the FRW Universe in \cite{Dynamical surface gravity AH}, with which the Cai--Kim  temperature satisfies $\displaystyle T_{\text  A}=\frac{\kappaup}{2\pi}$ at the horizon  $\Upsilon=\Upsilon_{\text{A}}$ and thus absorbs the Hayward temperature $\mathcal{T}_{\text  A}=\kappaup/(2\pi)$.

 With these considerations, we adopt the Cai--Kim $T_{\text{A}}$ for the absolute temperature of the cosmological apparent horizon.  This way, we believe that the temperature confusion is  solved as the Cai--Kim  $T_{\text{A}}$  is favored.

Furthermore, imagine a contracting Universe with $\dot{a}<0$ and $H<0$,  and one would have a marginally \textit{outer} trapped apparent horizon with $\theta_{(\ell)}=0$ and $\theta_{(n)}=2H<0 $ at $\Upsilon=\Upsilon_{\text{A}}$. Hence, whether  $\Upsilon=\Upsilon_{\text{A}}$ is outer or inner trapped only relies on the Hubble parameter to be negative or positive.
In Sec.~\ref{GR Induced metric of the apparent horizon} we have seen that the induced-metric signature of  $\Upsilon_{\text{A}}$ is independent of $H$, and neither will  the Hayward \{$\mathcal{T}_{\text{A}}$ , $\mathcal{T}_{\text{A}}^{(+)}$\}.
Also, Eqs.(\ref{Thermodynamic sign convention 1}) and (\ref{Thermodynamic sign convention 2}) clearly show that, the equality $-T_{\text{A}}dS_{\text{A}}=A_{\text{A}}\bm\psi_t=dE_{\text{A}}$ of the Cai--Kim $T_{\text{A}}$ validates for  either $H>0$ or $H<0$.
 Hence,   we further conclude that:\\

\textbf{Corollary 1} \textit{Neither the sign of the Hayward nor the Cai--Kim temperature is related to the inner or outer trappedness of the cosmological apparent horizon}.

\subsection{A quick note on the QCD ghost dark energy}

Among the various types of quantum chromodynamics (QCD)  ghost dark energy in  existent literature,  the following version was introduced in \cite{QCD ghost DE wrong} and further discussed in  \cite{QCD ghost DE wrong II},
\begin{equation}\label{QCD ghost DE}
\rho_\Lambda^{\text{(QCD)}}\,=\,\alpha\Upsilon_{\text{A}}^{-1} \left(1-\frac{\dot{\Upsilon}_{\text{A}}}{2H\Upsilon_{\text{A}}} \right)\,,
\end{equation}
where $\alpha$ is a positive constant with the dimension of $[energy]^3$. It is
based on the idea that the vacuum energy density is proportional to the temperature of the apparent horizon $\Upsilon_{\text{A}}$, which was chosen as the Hayward \{$\mathcal{T}_{\text{A}}$ , $\mathcal{T}_{\text{A}}^{(+)}$\} in \cite{QCD ghost DE wrong}. Following the discussion just above, we can see that Eq.(\ref{QCD ghost DE}) turns out to be problematic because $\rho_\Lambda^{\text{(QCD)}}$ is not positive definite, with $\rho_\Lambda^{\text{(QCD)}}\leq 0$ when the Universe is dominated by superrelativistic matter $1/3\leq w_m\;(\leq 1)$.
In fact, more viable forms of the QCD ghost dark energy can be found in e.g. \cite{QCD ghost DE correct}.


\section{The (generalized) second laws of thermodynamics}\label{The (generalized) second laws of thermodynamics}

Having studied the differential forms of the energy conservation and heat transfer and  distinguished the temperature of of the apparent horizon, we will proceed to investigate the  entropy evolution for the Universe.


\subsection{Positive heat out thermodynamic sign convention}\label{subsection Positive heat out thermodynamic sign convention}

As a corner stone for our formulation of the second laws and solution to the entropy confusion, we will match the thermodynamic sign convention encoded in
 the Cai--Kim Clausius equation $T_{\text{A}}dS_{\text{A}}=\delta Q_{\text{A}}=-A_{\text{A}}\bm\psi_t$.
Following Secs.~\ref{GR Unified first law of thermodynamics} and ~\ref{GR Clausius equation On the horizon}, we first check whether  the heat flow element $\delta Q_{\text{A}}$ and the isochoric energy differential $dE_{\text{A}0}=d(\rho_m V_{\text{A}0})$  take positive or negative values.  $\delta Q_{\text{A}}$ will be calculated by $T_{\text{A}}dS_{\text{A}}$, while $dE_{\text{A}}$ is to be evaluated independently
via $A_{\text{A}}\bm\psi_t=-A_{\text{A}} \big(1+w_m\big)\rho_m H\Upsilon_{\text{A}} dt$. Hence, in the isochoric process for an instantaneous apparent horizon $\Upsilon_{\text{A}0}$,
\begin{equation}\label{Thermodynamic sign convention 1}
T_{\text{A}}dS_{\text{A}}\,=\,  \frac{ \dot{\Upsilon}_{\text{A}}}{G}dt\,
=\, \frac{3}{2G} H \Upsilon_{\text{A}}\,\big(1 + w_m \big)dt\,,
\end{equation}
\begin{equation}\label{Thermodynamic sign convention 2}
dE \big|_{\Upsilon_{\text{A}0}}= -A_{\text{A}}\rho_m \big(1+w_m\big) H\Upsilon_{\text{A}} dt
 =- \frac{3}{2G}H\Upsilon_{\text{A}} \big(1+w_m\big)dt ,
\end{equation}
where Eqs. (\ref{GR Arho}) and (\ref{GR Friedmann Upsilon III}) have been used to replace $A_{\text{A}}\rho_m$ and $\dot{\Upsilon}_{\text{A}}$, respectively. For an expanding Universe ($H>0$),  this clearly shows that:
\begin{enumerate}
  \item[(1)] If the Universe is dominated by ordinary  matter or quintessence,  $-1< w_m\,(\leq 1)$, the internal energy is decreasing $dE_{\text{A}}=A_{\text{A}}\bm\psi_t<0$, with a positive Hubble energy flow $\delta Q_{\text{A}}=T_{\text{A}}dS_{\text{A}}>0$ going outside to the surroundings;
  \item[(2)] Under the dominance of the cosmological constant, $w_m=-1$ and $\{\rho_m, \Upsilon_{\text{A}} ,T_{\text{A}}  ,S_{\text{A}}  \}$ = constant;  the internal energy is unchanging, $dE_{\text{A}}=A_{\text{A}}\bm\psi_t=0$ and $\delta Q_{\text{A}}=T_{\text{A}}dS_{\text{A}}=0$;
  \item[(3)]  When the Universe  enters the phantom-dominated state,  $w_m<-1$, the internal energy increases $dE_{\text{A}}=A_{\text{A}}\bm\psi_t>0$ while $\delta Q_{\text{A}}=T_{\text{A}}dS_{\text{A}}<0$.
\end{enumerate}
Hence, based on the intuitive behaviors at the domain $-1<w_m\,(\leq1)$ for nonexotic matter, we set up the \emph{positive heat out} thermodynamic sign convention for the right hand side of  $dE_{\text{A}}=-\delta Q_{\text{A}}$.  That is to say, heat emitted by the system takes positive values ($\delta Q_{\text{A}}=\delta Q_{\text{A}}^{\text{out}}>0$), while heat absorbed by the system takes negative values. Obviously, this setup is totally consistent with the situations of $w_m\leq-1$.
Also, because of the counterintuitive behaviors under  phantom dominance, one should not take it for granted that,  for a spatially expanding Universe  the cosmic fluid would always flow out of the isochoric volume $V(\Upsilon=\Upsilon_0)$ with $dE=V_{\text{A}0}d\rho_m <0$.



\subsection{Positive heat out Gibbs equation}

In  existent papers, the cosmic entropy is generally studied independently of the first laws,
and the entropy $\widehat{S}_m$ of the cosmic energy-matter content (with temperature  $T_m$) is always determined by the  traditional Gibbs equation $dE=T_m d\widehat{S}_m-P_m dV$  (e.g. \cite{causal boundary entropy}-\cite{GSL scalar tensor chameleon}).
This way, $\dot{\widehat{S}}_m$ departs dramatically from the expected non-decreasing behaviors, so people  turn  to the generalized version of the second law for help, which works with the sum of $\widehat{S}_m$ and the geometric entropy of the cosmological apparent or event horizons.

This popular treatment is very problematic. In fact, the equation  $dE_m=T_m d\widehat{S}_m-P_m dV$ encodes  the ``positive heat in, positive work out''  convention for the physical entropy  $\widehat{S}_m$ and the heat transfer $T_m d\widehat{S}_m$. However,  as extensively discussed  just above, the geometric Bekenstein--Hawking entropy $S_{\text{A}}=A_{\text{A}}/4G$ for the cosmological apparent horizon
is compatible with the positive-heat-out convention. \emph{One cannot add the traditional positive-heat-in $\widehat{S}_m$ with the positive-heat-out  $S_{\text{A}}$}, and this  conflict\footnote{Note that there is no such conflict for black holes, because both the black-hole horizon entropy and the matter entropy are defined in the positive-heat-in convention.}  leads us to
adjust the Gibbs equation into
\begin{equation}\label{Gibbs Positive out}
dE_m \,=\,-T_m \,dS_m -P_mdV\,,
\end{equation}
where $S_m$ is defined in the  positive-heat-out convention favored by the Universe for consistency with the holographic-style gravitational equations  (\ref{GR Friedmann Upsilon}),  (\ref{GR Friedmann Upsilon II}) and   (\ref{GR Friedmann Upsilon IV}). This way, one can feel free and safe to superpose or compare the matter entropy $S_m$ and the horizon entropy \{$S_{\text{A}}$, etc.\}, and even more pleasantly, it turns out that this $S_m$ is very well behaved.

Moreover, note that  although the Gibbs equation is usually derived from a reversible process in a closed system (``controlled mass''), Eq.(\ref{Gibbs Positive out}) actually applies to either reversible or irreversible processes, and either closed or open systems, because it only contains state quantities which are independent of thermodynamic processes.

For the energy $E=M=\rho_m V$  in an arbitrary volume $V=\frac{4}{3}\pi \Upsilon^3=\frac{1}{3}A\Upsilon$, Eq.(\ref{Gibbs Positive out}) yields $T_m dS_m  =-d (\rho_m V  )-P_m dV=-V d\rho_m- (\rho_m + P_m )dV $, and thus
\begin{equation}\label{Gibbs 2nd law setup}
\begin{split}
T_m dS_m \,&=\;\;3H(\rho_m+P_m)V dt- \big(\rho_m + P_m\big)A d\Upsilon \\
&=\;\;\rho_m A (1+w_m)\big(H\Upsilon dt - d \Upsilon  \big)\,,
\end{split}
\end{equation}
where the continuity equation (\ref{GR continuity eqn}) has been used. Based on Eq.(\ref{Gibbs 2nd law setup}), we can analyze the entropy evolution $\dot{S}_m $ for the matter inside some special  radii such as the apparent and event horizons. Note that these regions  are generally open thermodynamic  systems   with the Hubble energy flow crossing the apparent and possibly the event horizons, so one should not \emph{a priori} anticipate  $\dot{S}_m \geq 0$; instead, we will look for the circumstances where  $\dot{S}_m \geq 0$ conditionally holds.




\subsection{The second law for the interior of the  apparent horizon}

For the matter inside the apparent horizon  $\Upsilon=\Upsilon_{\text{A}}(t)$,  Eq.(\ref{Gibbs 2nd law setup}) along with the holographic-style dynamical equations (\ref{GR Arho}) and (\ref{GR Friedmann Upsilon III}) yield
\begin{equation}\label{Gibbs Apparent}
\begin{split}
T_m dS_m^{(\text{A})} \,
&=\;\;\rho_m A_{\text{A}}(1+w_m)\big(H\Upsilon_{\text{A}} - \dot{\Upsilon}_{\text{A}}\big)dt\\
&=\;\;\frac{3}{2G}(1+w_m)H\Upsilon_{\text{A}}\Big(1 - \frac{3}{2}(1+w_m)\Big)dt\\
&=-\frac{9}{4G}H\Upsilon_{\text{A}}\big(w_m+1\big)\big(w_m + \frac{1}{3}\big)dt\,.
\end{split}
\end{equation}
Obviously  the second law of thermodynamics  $\dot{S}_m^{(\text{A})} \geq 0$ holds for  $-1\leq w_m\leq -1/3$. Moreover, recall that the spatial expansion of the generic FRW Unverse satisfies
\begin{equation}
\frac{\ddot{a}}{a}
\,=\,-\frac{4\pi G}{3}\big(1+3w_m \big)\rho_m\,,
\end{equation}
with $\ddot{a}> 0$ for $w_m<-1/3$. Hence,  within GR and the $\Lambda$CDM model, we have:\\

\textbf{Theorem 1} \textit{The physical entropy $S_m^{(\text{A})}$ inside the  cosmological apparent horizon satisfies $\dot{S}_m^{(\text{A})} \equiv 0$ when  $w_m=-1/3$ or under the dominance of the cosmological constant $w_m=-1$, while $\dot{S}_m^{(\text{A})} > 0$
 for the stage of  accelerated  expansion ($\ddot{a}>0$) dominated by quintessence $-1< w_m< -1/3$.}



\subsection{The second law for the interior of the  event and particle horizons}

Consider the future-pointed cosmological event horizon $\Upsilon_{\text{E}}\coloneqq a\int_t^\infty a^{-1}d\hat{t}$ which measures the distance that light signals will travel over the entire future history from $\hat{t}_0=t$. $\Upsilon_{\text{E}}$ satisfies
\begin{equation}
\dot{\Upsilon}_{\text{E}}=H\Upsilon_{\text{E}}-1\,,
\end{equation}
so for the cosmic fluid inside $\Upsilon_{\text{E}}$, Eq.(\ref{Gibbs 2nd law setup}) leads to
\begin{equation}\label{Second law EH w>-1}
\begin{split}
T_m dS_m^{(\text{E})} &=\rho_m A_{\text{E}}(1+w_m)\big(H\Upsilon_{\text{E}} - \dot{\Upsilon}_{\text{E}}\big)dt\\
&=\rho_m A_{\text{E}}(1+w_m)dt\,.
\end{split}
\end{equation}
Hence,  we are very happy to see that: \\

\textbf{Theorem 2} \textit{The physical entropy $S_m^{(\text{E})} $ inside the cosmological event horizon satisfies
$\dot{S}_m^{(\text{E})}  \equiv 0$  if
 the Universe is dominated by the cosmological constant $w_m=-1$, while $\dot{S}_m^{(\text{E})}  > 0$ for all nonexotic matter  $-1<w_m\,(\leq 1)$ above the phantom divide.}\\

The importance of this result can be best seen for a closed ($k=1$) Universe, when
the  event horizon   $\Upsilon_{\text{E}}$ has a  finite radius and bounds
 the entire spacetime. Then the physical entropy of the whole Universe is nondecreasing as long as the dominant energy condition holds $-1\leq w_m\, (\leq1)$.

Similarly for the past particle horizon $\Upsilon_{\text{P}}\coloneqq a\int_0^t a^{-1} d\hat{t}$ (e.g. \cite{Bak Cosmic holography, Faraoni, Four horizons two ways}), which supplements the event horizon $\Upsilon_{\text{E}}$ and measures the distance that light has already traveled from the beginning of time (or equivalently the most distant objects one could currently observe), it satisfies $\dot{\Upsilon}_{\text{P}}=H\Upsilon_{\text{P}}+1$ and thus Eq.(\ref{Gibbs 2nd law setup}) yields
\begin{equation}\label{Second law EH w>-1 PH}
\begin{split}
T_m dS_m^{(\text{P})} &=\rho_m A_{\text{P}}(1+w_m)\big(H\Upsilon_{\text{P}} - \dot{\Upsilon}_{\text{P}}\big)dt\\
&=-\rho_m A_{\text{P}}(1+w_m)dt\,.
\end{split}
\end{equation}
Besides $\dot{S}_m^{(\text{P})}\equiv0$ for $w_m=-1$,  $\dot{S}_m^{(\text{P})} <0$  always holds at the  domain $-1<w_m\,(\leq 1)$, which means that the physical entropy is always decreasing when we trace back to the earlier age for the younger Universe that has a larger particle horizon radius $\Upsilon_{\text{P}}$ or horizon area $A_{\text{P}}$.

Note that with the traditional Gibbs equation $dE_m =T_m d\widehat{S}_m-P_mdV$ where $\widehat{S}_m$ is defined in the positive-heat-in convention,  for the interiors of the future $\Upsilon_{\text{E}}$ and the past $\Upsilon_{\text{P}}$ one would always obtain
\begin{equation}\label{Second law EH w<-1}
\begin{split}
T_m \,d\widehat{S}_m^{(\text{E})} \,&=\,dE_m^{(\text{E})}  +P_mdV_{\text{E}}\,=-\rho_m A_{\text{E}}(1+w_m)dt\\
T_m \,d\widehat{S}_m^{(\text{P})} \,&=\,dE_m^{(\text{P})}  +P_mdV_{\text{P}}\:=\;\;\rho_m A_{\text{P}}(1+w_m)dt\,.
\end{split}
\end{equation}
It would imply that in the future $\dot{\widehat{S}}_m^{(\text{E})} >0$ would never be realized and a younger Universe (larger $A_{\text{P}}$) would however carry a larger internal entropy $\widehat{S}_m^{(\text{P})}$,  unless the Universe were in an exotically phantom-dominated ($w_m<-1$) state in her history. We believe that Eqs.(\ref{Second law EH w>-1}, \ref{Second law EH w>-1 PH}) provide a more reasonable description for the cosmic entropy evolution than  Eq.(\ref{Second law EH w<-1}),  regard this result as a support to  the positive-heat-out Gibbs equation (\ref{Gibbs Positive out}), and argue that Eqs.(\ref{Gibbs Positive out}), (\ref{Second law EH w>-1}) and  (\ref{Second law EH w>-1 PH}) have solved the cosmological \textit{entropy confusion} caused by Eq.(\ref{Second law EH w<-1}) in traditional studies.


\subsection{GSL for the apparent-horizon system}\label{Generalized second law for the apparent-horizon system}

Historically, to rescue the disastrous result of the traditional Eq.(\ref{Second law EH w<-1}), the \textit{generalized} second law (GSL) for the thermodynamics of the Universe was
developed, which  adds up the geometrically defined entropy of the cosmological boundaries  (mainly $S_{\text{A}}, S_{\text{E}}$) to the physical entropy of the matter-energy content $S_m$, aiming to make the total entropy nondecreasing under certain conditions. This idea is inspired by the GSL of black-hole  thermodynamics \cite{GSLT BH}, for which Bekenstein postulated that  the black-hole horizon entropy plus the external matter entropy never decrease (for a thermodynamic closed system).

Eq.(\ref{Second law EH w>-1}) clearly indicates that the second law $\dot{S}_m\geq 0$ is well respected in our formulation, but for completeness we will still re-investigate the GSLs.
For the simple open system consisting of the cosmological apparent horizon $\Upsilon_{\text{A}}$ and its interior, Eqs.(\ref{GR Hawking-Bekenstein entropy}) and (\ref{Gibbs Apparent}) yields
\begin{eqnarray}
\dot{S}_m^{\text{(A)}} +\dot{S}_{\text{A}}= &&-\frac{1}{T_m }\frac{9}{4G}H\Upsilon_{\text{A}}
\big(w_m+1\big)\big(w_m + \frac{1}{3}\big)+\frac{2\pi\Upsilon_{\text{A}}\dot{\Upsilon}_{\text{A}} }{G}\nonumber\\
=&&-\frac{1}{T_m } \frac{9}{4G}H\Upsilon_{\text{A}}
\big(w_m+1\big)\big(w_m + \frac{1}{3}\big)\label{GSLT condition 1} \\
&&+\frac{1}{T_{\text{A}}}\frac{3}{2G}H\Upsilon_{\text{A}}\big(w_m+1\big).\nonumber
\end{eqnarray}
In existing papers it is generally assumed that  the apparent horizon would be in thermal equilibrium with the matter content and thus $T_{\text{A}}=T_m$ \cite{GSL Interacting holographic DE, GSL MG Akbar, GSL Gauss-Bonnet braneworld, GSL Horava-Lifshitz, GSL scalar tensor chameleon, EventH GSLT Gauss Bonnet, GSLT f R interacting}, or occasionally less restrictively $T_m=bT_{\text{A}}$ ($b=$ constant) \cite{2nd law modified gravity, EventH GSL fR flat}. However, such assumptions are essentially  mathematical tricks to simplify Eq.(\ref{GSLT condition 1}), while physically they  are too
problematic, so we directly move ahead from Eq.(\ref{GSLT condition 1}) without
 any artificial speculations  relating $T_{\text{A}}$ and $T_m$.

The GSL $\dot{S}_m^{\text{(A)}} +\dot{S}_{\text{A}}\geq 0$ could hold when $\frac{1}{T_{\text{A}}}\frac{3}{2G}H\Upsilon_{\text{A}}\big(w_m+1\big)\geq \frac{1}{T_m } \frac{9}{4G}H\Upsilon_{\text{A}}
\big(w_m+1\big)\big(w_m + \frac{1}{3}\big) $, and with $\{H,\Upsilon_{\text{A}}, T_{\text{A}}, T_m \}>0$ it leads to
\begin{equation}\label{GSLT condition 2}
\begin{split}
\big(w_m+1\big) \Bigg( \frac{T_m }{T_{\text{A}}}-\frac{3}{2}
\big(w_m + \frac{1}{3}\big)  \Bigg)\,\geq\, 0\,,
\end{split}
\end{equation}
or equivalently $\big(w_m+1\big) \big(T_m -\frac{3}{2}
(w_m + \frac{1}{3})T_{\text{A}}  \big)\geq 0$.  Hence,  for the apparent-horizon system the GSL trivially validates with $\dot{S}_m +\dot{S}_{\text{A}}\equiv 0$ under the dominance of the cosmological constant  $w_m=-1$, and:
\begin{enumerate}
  \item[(1)]   For  $-1<w_m< -1/3$ which corresponds to an accelerated  Universe dominated by quintessence,
$\dot{S}_m +\dot{S}_{\text{A}} > 0$ always holds, because $T_m/ T_{\text{A}} >0$ and  $\frac{3}{2}\big(w_m + \frac{1}{3} \big)<0$ [or because  both $\dot{S}_m>0$ and $\dot{S}_{\text{A}} > 0$];

  \item[(2)] For $ -1/3\leq w_m\,(\leq 1)$  which corresponds to ordinary-matter dominance respecting the strong energy condition $\rho_m+3P_m\geq 0$ \cite{Hawking Ellis},   the GSL $\dot{S}_m +\dot{S}_{\text{A}} \geq 0$ conditionally holds when
  \begin{equation}
\frac{T_m }{T_{\text{A}}}\,\geq\,\frac{3}{2}
\big(w_m + \frac{1}{3}\big)\,;
 \end{equation}
  \item[(3)] For the phantom domain $w_m<-1$, the GSL never validates because it requires $T_m/T_{\text{A}} \leq \frac{3}{2}
 (w_m + \frac{1}{3} )<0$
which  violates the  the third law of thermodynamics.
\end{enumerate}


\subsection{GSL for the event-horizon system}\label{Generalized second law for the event-horizon system}

Now consider the system made up of the cosmological  event horizon   and its interior.
Unlike the apparent horizon, the entropy  $S_{\text{E}}$ and temperature $T_{\text{E}}$ of the event horizon  $\Upsilon_{\text{E}}$ are  unknown yet;
one should not take it for granted that $\Upsilon_{\text{E}}$ would still carry the Bekenstein--Hawking entropy $S_{\text{E}}=A_{\text{E}}/4G$ and further assume a Cai--Kim temperature $T_{\text{E}}={1}/(2\pi \Upsilon_{\text{E}})$ to it.

Considering that $S_{\text{E}}$ would reflect the amount of Hubble-flow energy crossing an instantaneous event horizon $\Upsilon_{\text{E}}=\Upsilon_{\text{E}0}$, it is still safe to make use of the unified first law  Eq.(\ref{GR dE inside t Upsilon wu}) and thus
 \begin{equation}\label{GSL EH correct 1}
T_{\text{E}} dS_{\text{E}} \,=\,\delta Q_{\text{E}}\,=- dE\,\big|_{\Upsilon_{\text{E}0}}=\,A_{\text{E}}\,\big(1+w_m\big)\rho_m \,H\Upsilon_{\text{E}}\,dt\,.
\end{equation}
Hence for the event horizon system we have
\begin{eqnarray}\label{GSL EH correct 1.6}
\dot{S}_m^{\text{(E)}} +\dot{S}_{\text{E}} &=& \frac{1}{T_m}  (1+w_m)\rho_m A_{\text{E}}
+\frac{1}{T_{\text{E}}}A_{\text{E}} \big(1+w_m\big)\rho_m  H\Upsilon_{\text{E}}  \nonumber \\
&=&   \rho_m A_{\text{E}} \big(1+w_m\big) \Big(\frac{1}{T_m} +\frac{1}{T_{\text{E}}}\frac{\Upsilon_{\text{E}}}{\Upsilon_{\text{H}}} \Big) ,
\end{eqnarray}
where $\Upsilon_H\coloneqq 1/H$ refers to the radius of the  Hubble horizon \cite{FRW studies, Faraoni}, an  auxiliary scale where the recession speed would reach that of light ($c=1$ in our units) by Hubble's law, and it is more instructive to write $H$ as $1/\Upsilon_{\text{H}}$ when compared with $\Upsilon_{\text{E}}$ and $\Upsilon_{\text{A}}$.
Since $\{T_{\text{A}}, T_{\text{E}}, H, \Upsilon_{\text{E}}\}>0$,  we pleasantly conclude from  Eq.(\ref{GSL EH correct 1.6}) without any unnatural assumption on $\{T_m, T_{\text{E}}\}$ that:\\

\textbf{Theorem 3} \textit{The GSL  $\dot{S}_m^{\text{(E)}} +\dot{S}_{\text{E}}\geq 0$  for the event horizon system always holds for an expanding Universe dominated by nonexotic matter $-1\leq w_m\,(\leq 1)$}.\\

Note that Mazumder and Chakraborty have discussed GSLs for the event-horizon system in various dark-energy (and modified-gravity) models in \cite{Event H entropy correct I, Event H entropy correct II}, where $S_{\text{E}}$ is calculated by the unified first law and the importance of $w_m$ is fully realized, although it is the weak rather than the dominant energy condition that is emphasized therein and the possibility of a Bekenstein--Hawking entropy for  $\Upsilon_{\text{E}}$  is not analyzed.

So far we have seen that though the apparent horizon $\Upsilon_{\text{A}}$ is more compatible with the unified first law and the Clausius equation,  the
second law is better respected by the cosmic fluid inside the event horizon  $\Upsilon_{\text{E}}$ -- this is because $\Upsilon_{\text{E}}$ better captures the philosophical concept of ``the whole Universe''. For both horizons $\Upsilon_{\text{A}}$ and   $\Upsilon_{\text{E}}$, the second law is better formulated than the GSL. Moreover, from the standpoint of the second laws and the GSLs,
 the phantom ($w_m<-1$) dark energy is definitely less favored than the cosmological constant ($w_m=-1$) and  the quintessence ($-1< w_m< -1/3$).


\subsection{Bekenstein--Hawking entropy and Cai--Kim temperature for the event horizon?}\label{Bekenstein-Hawking entropy and Cai--Kim temperature for the event horizon}

The entropy of the event horizon  $\Upsilon_{\text{E}}$ has just been calculated from the unified fist law. Now let's return to the question: Can the Bekenstein--Hawking entropy and/or the Cai--Kim temperature be applied to $\Upsilon_{\text{E}}$? With the assumption
$S_{\text{E}}=A_{\text{E}}/4G$, Eq.(\ref{GSL EH correct 1}) yields
\begin{equation}
\hspace{-2mm}T_{\text{E}} \frac{2\pi\Upsilon_{\text{E}}\dot{\Upsilon}_{\text{E}} }{G}
 = T_{\text{E}} \frac{2\pi\Upsilon_{\text{E}}\big(H\Upsilon_{\text{E}}-1\big)}{G} =
A_{\text{E}} \big(1+w_m\big)\rho_m  H\Upsilon_{\text{E}} ,
\end{equation}
which further leads to
\begin{equation}\label{GSL EH correct 2}
\big(\Upsilon_{\text{E}}-\Upsilon_{\text{H}}\big)\,T_{\text{E}}\,=\,
\frac{G}{2\pi}\,\rho_m A_{\text{E}}\big(1+w_m\big)\,.
\end{equation}
An expanding FRW Universe always satisfies $\Upsilon_{\text{E}}\geq\Upsilon_{\text{H}}$, so
the third law of thermodynamics $T_{\text{E}}>0$ requires $-1\leq w_m\,(\leq 1)$; also,
$\Upsilon_{\text{E}}=\Upsilon_{\text{H}}$ when $w_m=-1$ and $T_{\text{E}}$ becomes unspecifiable from Eq.(\ref{GSL EH correct 2}). Moreover,
if $\Upsilon_{\text{E}}=\Upsilon_{\text{H}}$, then $a\int_t^\infty a^{-1}d\hat{t}=\frac{a}{\dot a}$, thus
\begin{equation}\label{GSL EH correct 3}
\dot{a}\int_t^\infty a^{-1}d\hat{t}=1\quad\Rightarrow\quad \frac{\ddot a}{\dot a}-\frac{\dot a}{a}=0
\quad\Rightarrow\quad a\ddot a={\dot a}^2\,,
\end{equation}
where we have taken the time derivative of the left-most integral expression. In the meantime, when $w_m=-1$ we have
\begin{equation}\label{GSL EH correct 4}
\frac{{\dot a}^2+k}{a^2}=\frac{8}{3}\pi G\rho_m\;\;\,,\;\;
\frac{\ddot a}{a}=\frac{8}{3}\pi G\rho_m
\;\;\Rightarrow\;\; a\ddot a= {\dot a}^2+k.
\end{equation}
Comparison of Eqs.(\ref{GSL EH correct 3}) and (\ref{GSL EH correct 4}) shows that in addition to $w_m=-1$,  $\Upsilon_{\text{E}}=\Upsilon_{\text{H}}$ also requires $k=0$; note that in case of the flat Universe, the apparent and the Hubble horizons coincide,  $\Upsilon_{\text{A}}=\Upsilon_{\text{H}}$, so $T_{\text{E}}$=$T_{\text{A}}$ which remedies the failure of Eq.(\ref{GSL EH correct 2}) at $w_m=-1$. Hence,\\

\textbf{Corollary 2} \textit{The validation of a Bekenstein--Hawking entropy on the cosmological event horizon requires that (i) the scale factor $a(t)$ satisfies the constraint Eq.(\ref{GSL EH correct 2}), (ii) the dominant energy condition always holds, (iii) the event and Hubble horizons would coincide and the spatial curvature vanishes under the dominance of the cosmological constant.}\\

\noindent If one further assumes a Cai--Kim-like $T_{\text{E}}={1}/(2\pi \Upsilon_{\text{E}})$ for the event horizon, Eq.(\ref{GSL EH correct 2}) would tell us that
\begin{equation}\label{GSL EH correct 5}
G\rho_m A_{\text{E}}\,\big(1+w_m\big)+\frac{\Upsilon_{\text{H}}}{\Upsilon_{\text{E}}}\,=\,1\,.
\end{equation}
Does this constraint always hold? Since $\Upsilon_{\text{E}}\geq \Upsilon_{\text{A}}$, thus $\rho_m A_{\text{E}}\geq \rho_m A_{\text{A}}=\frac{3}{2G}$, with which Eq.(\ref{GSL EH correct 5}) yields
\begin{equation}
\frac{3}{2}\big(1+w_m\big)+\frac{\Upsilon_{\text{H}}}{\Upsilon_{\text{E}}} \,\leq \,1\,.
\end{equation}
This result can be rearranged into
\begin{equation}
w_m\,\leq-\frac{1}{3}-\frac{\Upsilon_{\text{H}}}{\Upsilon_{\text{E}}}\,<-\frac{1}{3}\,,
\end{equation}
which, together with the requirement  $-1\leq w_m\,(\leq 1)$ from Eq.(\ref{GSL EH correct 2}) for a generic positive $T_{\text{E}}$, give rise to the condition $-1\leq w_m<-1/3$. Hence,\\

\textbf{Corollary 3} \textit{In addition to a Bekenstein--Hawking entropy, the validation of a Cai--Kim temperature on the cosmological event horizon further requires the scale factor to satisfy Eq.(\ref{GSL EH correct 5}), and restricts the FRW Universe to be dominated by the cosmological constant $w_m=-1$ or quintessence $-1<w_m<-1/3$.}\\

\noindent Similar conditions hold for the past particle horizon as well.  \cite{Four horizons two ways}
has derived the GSL inequalities for the Hubble-, apparent-, particle- and event-horizon systems with the logamediate and intermediate scale factors by both the first law and the Bekenstein--Hawking formula, in which one could clearly observe that these two methods yield different results in the case of the event (and particle, Hubble) horizons.

Based on these considerations we argue that for consistency with the cosmic gravitational dynamics, the geometrically defined $A/4G$ only unconditionally holds on the apparent horizon $\Upsilon_{\text{A}}$, which does not support the belief that the Bekenstein--Hawking entropy could validate for all horizons in GR (e.g. \cite{causal boundary entropy, Faraoni}).


\section{Gravitational thermodynamics in ordinary modified gravities}\label{Gravitational thermodynamics in ordinary modified gravities}

For the  $\Lambda$CDM  Universe within GR, we have re-studied the first and second laws of thermodynamics by requiring the consistency with the holographic-style dynamical equations (\ref{GR Friedmann Upsilon}), (\ref{GR Friedmann Upsilon II}) and (\ref{GR Friedmann Upsilon III}),
which provides possible solutions to the long-standing temperature and entropy confusions. Following  the clarification of the Cai--Kim temperature and the positive-heat-out sign convention, we will take this opportunity to extend the whole framework of gravitational thermodynamics to modified and  alternative  theories of  relativistic  gravity \cite{Dark Energy II, AA Tian-Booth Paper}; also, this is partly a continuation of our earlier work in \cite{AA our first law}  where a unified formulation has been developed to derive the cosmological dynamical equations in modified gravities from (non)equilibrium thermodynamics.

For the generic Lagrangian density $\mathscr{L}_{\text{total}}=\mathscr{L}_G(R, R_{\mu\nu}R^{\mu\nu},\mathcal{R}_{\,i}\,,
\vartheta\,, \nabla_\mu\vartheta\nabla^\mu\vartheta\,,\cdots\big)+16\pi G\mathscr{L}_m $, where $\mathcal{R}_i=\mathcal{R}_i\,\big(g_{\alpha\beta}\,,R_{\mu\alpha\nu\beta}\,,\nabla_\gamma R_{\mu\alpha\nu\beta}\,,\ldots\big)$ refers to a generic Riemannian invariant beyond the Ricci scalar and $\vartheta$ denotes a scalarial extra degree of freedom unabsorbed by $\mathscr{L}_m$\,, the field equation reads
\begin{equation}\label{FieldEqnGRForm 0}
H_{\mu\nu}=8\pi G T_{\mu\nu}^{(m)}\;\;\mbox{with}\;\;
H_{\mu\nu}\,\coloneqq\, \frac{1}{\sqrt{-g}} \frac{\delta\, \Big(\!\!\sqrt{-g}\,\mathscr{L}_G \Big)}{\delta g^{\mu\nu}}\,,
\end{equation}
where total-derivative/boundary terms should be removed in the derivation of $H_{\mu\nu}$. In the spirit of  reconstructing the effective dark energy \cite{Reconstructing dark energy},
Eq.(\ref{FieldEqnGRForm 0}) can be \emph{intrinsically} recast into a compact GR form by isolating the $R_{\mu\nu}$ in $H_{\mu\nu}$:
\begin{equation}\label{FieldEqnGRForm}
\begin{split}
G_{\mu\nu}\equiv
R_{\mu\nu} -\frac{1}{2}Rg_{\mu\nu} = &8\pi G_{\text{eff}} T_{\mu\nu}^{\text{(eff)}}\;\;\text{with}\\
H_{\mu\nu}=\frac{G}{G_{\text{eff}}}G_{\mu\nu}-&8\pi G T_{\mu\nu}^{\text{(MG)}}\,,
\end{split}
\end{equation}
where $T_{\mu\nu}^{\text{(eff)}} = T_{\mu\nu}^{(m)}+T_{\mu\nu}^{\text{(MG)}}$, and all terms
beyond GR have been packed into $T_{\mu\nu}^{\text{(MG)}}$ and $G_{\text{eff}}$.
Here  $T_{\mu\nu}^{\text{(MG)}}$ collects the modified-gravity nonlinear and higher-order effects,
while   $G_{\text{eff}}$ denotes the effective gravitational coupling strength which  can be
directly recognized from the coefficient of the matter tensor $T_{\mu\nu}^{(m)}$ -- for example,
as will be shown in Sec.\ref{subsection Applications}, we have $G_{\text{eff}}=G/f_R$ for $f(R)$, $G_{\text{eff}}=GE(\phi)/F(\phi)$ for scalar-tensor-chameleon,
$G_{\text{eff}}=G/\phi$ for Brans-Dicke,  $G_{\text{eff}}=G/(1+2aR)$ for quadratic,
and $G_{\text{eff}}=G$ for dynamical Chern-Simons gravities.  Moreover, $T_{\mu\nu}^{\text{(eff)}}$
is assumed to be an effective perfect-fluid content,
\begin{equation}\label{Effective SEM Perfect fluid}
\begin{split}
&T^{\mu\,\text{(eff)}}_{\;\;\nu}=\text{diag}\left[-\rho_{\text{eff}},P_{\text{eff}},P_{\text{eff}},P_{\text{eff}}\right]\\
&\text{with}\quad P_{\text{eff}}/\rho_{\text{eff}}\eqqcolon w_{\text{eff}},
\end{split}
\end{equation}
along with $\rho_{\text{eff}}=\rho_m +\rho_{\text{(MG)}} $ and $P_{\text{eff}}=P_m+P_{\text{(MG)}}$.

Modified gravities aim to explain the cosmic acceleration without dark-energy components, so in this section we will assume the physical matter to respect the null, weak, strong and dominant energy conditions \cite{Hawking Ellis}, which yield $\rho_m>0$ and $-1/3\leq w_m\leq 1$. This way, the quintessence ($-1<w_m<-1/3$), the cosmological constant ($w_m=-1$) and the most exotic phantom ($w_m<-1$) are ruled out.


\subsection{Holographic-style dynamical equations in modified gravities}\label{section From Friedmann equations to the unified first law}

Substituting the FRW metric
Eq.(\ref{FRW metric I}) and the effective cosmic fluid Eq.(\ref{Effective SEM Perfect fluid}) into the field equation (\ref{FieldEqnGRForm}), one could obtain the modified Friedmann equations
\begin{equation}\label{MG Friedmann eqn 1st}
\begin{split}
H^2+\frac{k}{a^2}&= \frac{8\pi G_{\text{eff}}}{3}\rho_{\text{eff}} \quad\text{and}\\
\dot H-\frac{k}{a^2}
=-4\pi G_{\text{eff}} \big(1&+w_{\text{eff}}\big) \rho_{\text{eff}}=-4\pi G_{\text{eff}} h_{\text{eff}}\\
\text{or}\quad2\dot H+3H^2+\frac{k}{a^2}&= -8\pi G_{\text{eff}} P_{\text{eff}} ,
\end{split}
\end{equation}
where $h_{\text{eff}} \coloneqq \big(1+w_{\text{eff}}\big)\rho_{\text{eff}}$ denotes the effective enthalpy density. With Eqs. (\ref{Horizon location}) and (\ref{dot Upsilon}), substituting the apparent-horizon radius $\Upsilon_{\text{A}}$ and its kinematic time-derivative $\dot{\Upsilon}_{\text{A}}$  into  Eq.(\ref{MG Friedmann eqn 1st}),
 the Friedmann equations  can be rewritten into
\begin{eqnarray}
&&\Upsilon_{\text{A}}^{-2}= \frac{8\pi G_{\text{eff}}}{3} \rho_{\text{eff}}\label{MG Friedmann Upsilon}\\
\dot{\Upsilon}_{\text{A}}= &&  4\pi  H \Upsilon_{\text{A}}^3  G_{\text{eff}} \big(1+w_{\text{eff}}\big)\rho_{\text{eff}}\label{MG Friedmann Upsilon II}\\
= &&  \frac{3}{2}H\Upsilon_{\text{A}}\big( 1+w_{\text{eff}}\big)\label{MG Friedmann Upsilon III}\\
\Upsilon_{\text{A}}^{-3}\Big(\dot{\Upsilon}&&{}_{\text{A}} -\frac{3}{2}H \Upsilon_{\text{A}}  \Big)=  4\pi G_{\text{eff}}HP_{\text{eff}}\label{MG Friedmann Upsilon IV} ,
\end{eqnarray}
along with $A_{\text{A}}\rho_{\text{eff}}= \frac{3}{2G_{\text{eff}}}$. Similar to Eqs.(\ref{GR Friedmann Upsilon})-(\ref{GR Friedmann Upsilon IV}) for $\Lambda$CDM of GR, Eqs.(\ref{MG Friedmann Upsilon})-(\ref{MG Friedmann Upsilon IV}) constitute the full set of FRW holographic-style gravitational equations for modified gravities of the form Eq.(\ref{FieldEqnGRForm}).


\subsection{Unified first law of nonequilibrium thermodynamics}\label{Unified first law of thermodynamics}

Following our previous work \cite{AA our first law}, to geometrically reconstruct the effective total internal energy $E_{\text{eff}}$, one just needs to replace Newton's constant $G$ by $G_{\text{eff}}$ in the standard Misner-Sharp or Hawking mass used in Sec.~\ref{GR Unified first law of thermodynamics}, which yields
\begin{equation}\label{MG mass}
E_{\text{eff}}
 = \frac{1}{2G_{\text{eff}}}\frac{\Upsilon^3}{\Upsilon_{\text{A}}^2} .
\end{equation}
The total derivative of $E_{\text{eff}}=E_{\text{eff}}(t,r)$ along with the  holographic-style dynamical equations (\ref{MG Friedmann Upsilon}), (\ref{MG Friedmann Upsilon II}) and (\ref{MG Friedmann Upsilon IV}) yield
\begin{eqnarray}
dE_{\text{eff}} =&&-\frac{1}{G_{\text{eff}}}\frac{ \Upsilon^3}{\Upsilon_{\text{A}}^3} \Big( \dot{\Upsilon}_{\text{A}}-\frac{3}{2}H \Upsilon_{\text{A}}  \Big) dt
+\frac{3}{2G_{\text{eff}}} \frac{\Upsilon^2}{\Upsilon_{\text{A}}^2} adr \nonumber\\
&&-\frac{\dot{G}_{\text{eff}}}{2G_{\text{eff}}^2}\frac{\Upsilon^3}{\Upsilon_{\text{A}}^2} dt \label{dM inside t r}\\
=&& -A \Upsilon  H  P_{\text{eff}} dt+A \rho_{\text{eff}}  adr
-V\frac{\dot{G}_{\text{eff}}}{G_{\text{eff}}}   \rho_{\text{eff}}   dt \label{dE inside t r} .
\end{eqnarray}
By the replacement $adr=d\Upsilon-H\Upsilon dt$, Eqs.(\ref{dM inside t r}) and (\ref{dE inside t r}) can be recast into the $(t,\Upsilon)$ transverse coordinates as
\begin{eqnarray}
\hspace{-3.9mm}dE_{\text{eff}} &=&-\frac{\dot{\Upsilon}_{\text{A}}}{G_{\text{eff}}}\frac{\Upsilon^3}{\Upsilon_{\text{A}}^3} dt+\frac{3}{2G_{\text{eff}}} \frac{\Upsilon^2}{\Upsilon_{\text{A}}^2}  d\Upsilon-
\frac{\dot{G}_{\text{eff}}}{2G_{\text{eff}}^2}\frac{\Upsilon^3}{\Upsilon_{\text{A}}^2} dt \label{dM inside t Upsilon}\\
&=&-A \big(1+w_{\text{eff}}\big)\rho_{\text{eff}} H\Upsilon dt
+ A \rho_{\text{eff}} d\Upsilon-V\frac{\dot{G}_{\text{eff}}}{G_{\text{eff}}}   \rho_{\text{eff}}  dt. \label{dE inside t Upsilon}
\end{eqnarray}
Both Eqs.(\ref{dE inside t r}) and (\ref{dE inside t Upsilon}) can be compactified  into the thermodynamic equation
\begin{eqnarray}\label{Unified first law}
dE_{\text{eff}} =  A\Psi+\mathcal{W} dV+\mathcal {E} ,
\end{eqnarray}
where $\mathcal{W}$ and $\Psi$ respectively
refer  to the effective work density and the effective energy supply covector,
\begin{eqnarray}
\mathcal{W} = \frac{1}{2} \big(1-&&w_{\text{eff}}\big) \rho_{\text{eff}}  ,\\
\Psi =-\frac{1}{2}  \big(1+w_{\text{eff}}\big)\rho_{\text{eff}} H\Upsilon&& dt
+ \frac{1}{2}  \big(1+w_{\text{eff}}\big)\rho_{\text{eff}} adr \nonumber \\
= \;\; - \big(1+w_{\text{eff}}\big)\rho_{\text{eff}} H\Upsilon&& dt
+ \frac{1}{2}  \big(1+w_{\text{eff}}\big)\rho_{\text{eff}} d\Upsilon,\label{psi flux density}
\end{eqnarray}
and similar to Sec.~\ref{GR Unified first law of thermodynamics}, $\mathcal{W}$ and $\Psi$ can trace  back to the Hayward-type invariants $\mathcal{W} \coloneqq -\frac{1}{2} T^{\alpha\beta}_{\text{(eff)}} h_{\alpha\beta}$ and $\Psi_\alpha \coloneqq T_{\alpha \text{(eff)}}^{\;\; \beta} \partial_\beta \Upsilon+\mathcal{W} \partial_\alpha \Upsilon$ under spherical symmetry. The $\mathcal{E}$ in  Eq.(\ref{Unified first law}) is an extensive energy term
\begin{equation}\label{Dissipation energy density}
\mathcal {E} \coloneqq  -V\frac{\dot{G}_{\text{eff}}}{G_{\text{eff}}}   \rho_{\text{eff}}\; dt .
\end{equation}
As will be shown in the next subsection,  $\mathcal{E}$ contributes to the irreversible extra entropy production, so we regard Eq.(\ref{Unified first law}) as the unified first law of \textit{nonequilibrium} thermodynamics \cite{AA our first law}, which is an extension of the equilibrium version Eq.(\ref{GR Unified first law}) in GR. Moreover,  it follows from the contracted Bianchi identities and  Eq.(\ref{FieldEqnGRForm}) that  $\nabla_\mu G^{\mu}_{\;\; \nu}=0=8\pi\nabla_\mu \big(G_{\text{eff}} T^{\mu \text{(eff)}}_{\;\;\nu} \big)$,   and for the FRW metric Eq.(\ref{FRW metric I}) it leads to
\begin{equation}\label{Generalized continuity eqn}
 \dot{\rho}_{\text{eff}}
+ 3 H \big(\rho_{\text{eff}}+P_{\text{eff}}  \big)    =  \frac{\dot{\mathcal{E}}}{V}  = -
\frac{\dot{G}_{\text{eff}}}{G_{\text{eff}}}   \rho_{\text{eff}} ,
\end{equation}
so $\mathcal{E}$ also shows up in the generalized continuity equation as a density  dissipation effect.


\subsection{Nonequilibrium Clausius equation on the horizon}\label{Nonequilibrium Clausius equation On the horizon}

The  holographic-style dynamical equation  (\ref{MG Friedmann Upsilon II}) can be slightly rearranged into
$\frac{ \dot{\Upsilon}_{\text{A}}}{G_{\text{eff}}} dt=A_{\text{A}} (1+w_{\text{eff}})\rho_{\text{eff}} H\Upsilon_{\text{A}}dt$,
so we have
\begin{equation}\label{MG On horizon step I}
\begin{split}
&\frac{1}{2 \pi \Upsilon_{\text{A}}}\cdot 2 \pi \Upsilon_{\text{A}}\left(\frac{ \dot{\Upsilon}_{\text{A}}}{G_{\text{eff}}} dt - \frac{1}{2} \Upsilon_{\text{A}} \frac{\dot{G}_{\text{eff}}}{G_{\text{eff}}^2}dt\right)+\\
&\frac{1}{2 \pi \Upsilon_{\text{A}}}\cdot 2 \pi \Upsilon_{\text{A}}\left( \frac{1}{2}\Upsilon_{\text{A}}\frac{\dot{G}_{\text{eff}}}{G_{\text{eff}}^2}   dt
+V_{\text{A}}\frac{\dot{G}_{\text{eff}}}{G_{\text{eff}}}   \rho_{\text{eff}}  dt\right)\\
&=A_{\text{A}} \big(1+w_{\text{eff}}\big) \rho_{\text{eff}} H\Upsilon_{\text{A}} dt
+V_{\text{A}}\frac{\dot{G}_{\text{eff}}}{G_{\text{eff}}}   \rho_{\text{eff}}  dt .
\end{split}
\end{equation}
It can be formally compactified into  the thermodynamic relation
\begin{equation}\label{MG TdS nonequi}
\begin{split}
T_{\text{A}}\left(dS_{\text{A}}+d_p S^{(\text{A})}\right)=-  \big(A_{\text{A}}\Psi_t+\mathcal{E}_{\text{A}}\big)=-dE_{\text{eff}}^{\text{A}}\Big|_{d\Upsilon=0},
\end{split}
\end{equation}
where $\Psi_t$ is just the $t$-component of the covector $\Psi$ in Eq.(\ref{psi flux density}), $\mathcal{E}_{\text{A}}$ is the energy dissipation term Eq.(\ref{Dissipation energy density}) evaluated at $\Upsilon_{\text{A}}$, and $T_{\text{A}}=\frac{1}{2 \pi \Upsilon_{\text{A}}}$ denotes the Cai--Kim temperature on  $\Upsilon_{\text{A}}$.
Here  $S_{\text{A}}$ refers to the geometrically defined Wald entropy \cite{Wald entropy} for the dynamical apparent horizon,
\begin{equation}\label{entropy Wald-Kodama}
S_{\text{A}} \;=\;\frac{\pi \Upsilon_{\text{A}}^2 }{G_{\text{eff}}}\;=\;\frac{A_{\text{A}}}{4G_{\text{eff}}} =  \int \frac{dA_{\text{A}}}{4G_{\text{eff}}} ,
\end{equation}
where  $S_{\text{A}}$ takes such a compact form due to $\Upsilon_{\text{A}}=\Upsilon_{\text{A}}(t)$ and $G_{\text{eff}}=G_{\text{eff}}(t)$ under the maximal spatial symmetry of the Universe, while
$d_pS^{(\text{A})} $ represents the irreversible entropy production within $\Upsilon_{\text{A}}$
\begin{equation}\label{MG dpS}
\begin{split}
d_pS^{(\text{A})}  &=  2 \pi \Upsilon_{\text{A}}\left( \frac{1}{2}\Upsilon_{\text{A}}\frac{\dot{G}_{\text{eff}}}{G_{\text{eff}}^2}   dt
+V_{\text{A}}\frac{\dot{G}_{\text{eff}}}{G_{\text{eff}}}   \rho_{\text{eff}}  dt\right)\\
&=  2 \pi \Upsilon_{\text{A}}^2 \frac{\dot{G}_{\text{eff}}}{G_{\text{eff}}^2}   dt ,\\
\end{split}
\end{equation}
where we have applied the following  replacement
\begin{equation}\label{MG horizon dissipation replacement}
\frac{1}{2}\Upsilon_{\text{A}}\frac{\dot{G}_{\text{eff}}}{G_{\text{eff}}^2}\;=\;
V_{\text{A}}\frac{\dot{G}_{\text{eff}}}{G_{\text{eff}}}   \rho_{\text{eff}} ,
\end{equation}
whose validity is guaranteed by Eq.(\ref{MG Friedmann Upsilon}).
Due to the extra entropy production element $d_pS^{(\text{A})}$, we regard Eq.(\ref{MG TdS nonequi}) as the \emph{nonequilibrium} Clausius equation,  which depicts the heat transfer plus the extensive energy dissipation for the isochoric process of an arbitrary instantaneous  $\Upsilon_{\text{A}}$. With the nonequilibrium unified first law Eq.(\ref{Unified first law}),  Eq.(\ref{MG TdS nonequi}) can be completed into the total energy differential
\begin{equation}
\begin{split}
dE_{\text{eff}}^{\text{A}}\;&=\;\;A_{\text{A}}\Psi_t  dt+A_{\text{A}}\left(\Psi_\Upsilon +\mathcal{W}\right) d\Upsilon_{\text{A}}+\mathcal{E}_{\text{A}}\\
&= -T_{\text{A}}\big(dS_{\text{A}}+d_pS^{(\text{A})}\big)+\rho_{\text{eff}} dV_{\text{A}} .
\end{split}
\end{equation}


\subsection{The second law for the interiors of the  apparent and  the  event horizons}

For the cosmic entropy evolution, the second law of thermodynamics should
still apply to the physical matter content $\{\rho_m ,P_m\}$
rather than the mathematically effective $\{\rho_{\text{eff}}, P_{\text{eff}}\}$. Under \emph{minimal} geometry-matter couplings,  the Noether compatible definition of $T_{\mu\nu}^{(m)}$ automatically guarantees  $\nabla^\mu T_{\mu\nu}^{(m)}=0$, so the total continuity equation (\ref{Generalized continuity eqn}) can be decomposed into the ordinary one for the physical matter and  the remaining part for the modified-gravity effect \cite{AA our first law}:
\begin{equation}\label{MG conservation}
\begin{split}
\dot{\rho}_m+3H (\rho_m+P_m  ) &= 0\\
\dot{\rho}_{\text{(MG)}}+3H \Big(\rho_{\text{(MG)}}+P_{\text{(MG)}}  \Big)
 &= -\frac{\dot{G}_{\text{eff}} }{G_{\text{eff}} } \Big(\rho_m+\rho_{\text{(MG)}}  \Big) .
\end{split}
\end{equation}
For the physical energy $E_m=\rho_m V=E_{\text{eff}}-\rho_{\text{(MG)}} V$ within an arbitrary volume, the positive-heat-out Gibbs equation (\ref{Gibbs Positive out}) still yields $T_m dS_m =-d (\rho_m V  )-P_m dV=-V d\rho_m- (\rho_m + P_m )dV $, which together with Eq.(\ref{MG conservation}) leads to
\begin{equation}\label{MG Gibbs 2nd law setup}
\begin{split}
T_m dS_m  &=\;\;3H(\rho_m+P_m)V dt- \big(\rho_m + P_m\big)A d\Upsilon \\
&=\;\;\rho_m A (1+w_m)\big(H\Upsilon dt - d \Upsilon  \big) .
\end{split}
\end{equation}
Hence, for the physical entropy $S_m^{\text{(A)}}$ inside the apparent horizon  $\Upsilon_{\text{A}}(t)$, Eq.(\ref{MG Gibbs 2nd law setup}) and the holographic-style dynamical equation (\ref{MG Friedmann Upsilon III}) yield
\begin{equation}\label{MG Gibbs AppH}
\begin{split}
T_m  dS_m^{\text{(A)}}
&= \;\;\rho_m A_{\text{A}}\big( 1+w_m \big)  \big(
\Upsilon_{\text{A}} H- \dot{\Upsilon}_{\text{A}}   \big)  dt\\
&= -\frac{3}{2}\rho_m A_{\text{A}}\big( 1+w_m \big) H\Upsilon_{\text{A}}\big(\frac{1}{3} + w_{\text{eff}}\big)dt\\
&= -\frac{9}{2}\rho_m V_{\text{A}} H \big( 1+w_m \big)\big(\frac{1}{3} + w_{\text{eff}}\big)dt .\\
\end{split}
\end{equation}
where $\rho_m A_{\text{A}}$ cannot be simplified by Eq.(\ref{GR Arho}) of GR. Recall that $ -1/3\leq w_m\leq 1$ in modified gravities, thus:\\

\textbf{Theorem 4} \textit{The physical entropy $S_m^{(\text{A})}$ inside the  cosmological apparent horizon satisfies $\dot{S}_m^{(\text{A})} \geq 0$ only when  $w_{\text{eff}} \leq -1/3$.}\\

\noindent Moreover, inside the event horizon  $\Upsilon_{\text{E}}(t)$,  Eq.(\ref{MG Gibbs 2nd law setup}) along with $\dot{\Upsilon}_{\text{E}}=H\Upsilon_{\text{E}}-1$ give rise to
\begin{equation}\label{MG EH SSL}
\begin{split}
T_m dS_m^{\text{(E)}}
&=\rho_m A_{\text{E}}(1+w_m)\big(H\Upsilon_{\text{E}} - \dot{\Upsilon}_{\text{E}}\big)dt\\
&=\rho_m A_{\text{E}}(1+w_m)dt .
\end{split}
\end{equation}
Hence,  for the FRW Universe governed by modified gravities and filled with ordinary matter $ -1/3\leq w_m\leq 1$:\\

\textbf{Theorem 5} \textit{The physical entropy $S_m^{(\text{E})}$ inside the  cosmological  event horizon always satisfies $\dot{S}_m^{(\text{E})} > 0$ regardless of the modified-gravity theories in use.}


\subsection{GSL for the apparent-horizon system}

Unlike the standard second law for the matter content $\{\rho_m ,P_m\}$, GSLs further involve the modified-gravity effects   $\{\rho_{(\text{MG})} ,P_{(\text{MG})}\}$ which influence the horizon entropy. Compared with the $\Lambda$CDM situation in Sec.~\ref{Generalized second law for the apparent-horizon system}, there are three types of entropy for the apparent-horizon system in modified gravities: the physical $S_m^{(\text{A})}$ for the internal matter content, the Wald  entropy $S_{\text{A}}$ of the horizon $\Upsilon_{\text{A}}$, and the nonequilibrium extensive entropy production. From Eqs.(\ref{MG TdS nonequi}) and (\ref{MG Gibbs AppH}), we have
\begin{eqnarray}
&&\dot{S}_m^{(\text{A})} +\dot{S}_{\text{A}}+\dot{S}_p^{(\text{A})}\nonumber\\
= &&-\frac{1}{T_m }\frac{3}{2}\rho_m A_{\text{A}}\big( 1+w_m \big) H\Upsilon_{\text{A}}\big(\frac{1}{3} + w_{\text{eff}}\big)\nonumber\\
&&+ \frac{ 2\pi\Upsilon_{\text{A}}\dot{\Upsilon}_{\text{A}}}{G_{\text{eff}}}+\pi\Upsilon_{\text{A}}^2\frac{\dot{G}_{\text{eff}}}{G_{\text{eff}}^2}\nonumber \\
=&& \frac{3}{2}\frac{\Upsilon_{\text{A}}}{\Upsilon_{\text{H}}}\left(-\frac{1}{T_m }\rho_m A_{\text{A}} \big( 1+w_m \big)\big(\frac{1}{3}+ w_{\text{eff}}\big)
+\frac{1}{T_{\text{A}}}\frac{1}{G_{\text{eff}}}\big(1 + w_{\text{eff}}\big)\right.\nonumber\\
&&\left.
+\frac{1}{T_{\text{A}}}\frac{1}{3H}\frac{\dot{G}_{\text{eff}}}{G_{\text{eff}}^2}  \right)\label{MG GSL AP 0} ,
\end{eqnarray}
where $\dot{S}_p^{(\text{A})}\coloneqq d_pS^{(\text{A})}/dt$,  $T_{\text{A}}=1/(2\pi\Upsilon_{\text{A}})$, and $\dot{\Upsilon}_{\text{A}}=\frac{3}{2}H\Upsilon_{\text{A}}\big(1 + w_{\text{eff}}\big)$. Generally  the GSL for the apparent-horizon system does not hold because the region $\Upsilon\leq \Upsilon_{\text{A}}$
only comprises a finite portion of the Universe and is thermodynamically open with the absolute
Hubble flow crossing $\Upsilon_{\text{A}}$. However, Eq.(\ref{MG GSL AP 0}) shows that
$\dot{S}_m^{(\text{A})} +\dot{S}_{\text{A}}+\dot{S}_p^{(\text{A})}\geq 0$ could validate when
\begin{equation}\label{MG GSLT}
\frac{T_m }{T_{\text{A}}} \left(\frac{1 + w_{\text{eff}}}{G_{\text{eff}}}
+\frac{1}{3H}\frac{\dot{G}_{\text{eff}}}{G_{\text{eff}}^2}\right)\geq
\rho_m A_{\text{A}} \big( 1+w_m \big)\big(\frac{1}{3} + w_{\text{eff}}\big),
\end{equation}
where  $A_{\text{A}}$ cannot be further replaced by $1/(\pi T_{\text{A}}^2)$ to nonlinearize $ T_{\text{A}}$ since $T_{\text{A}}$ is not an extensive quantity.
Specifically for equilibrium theories with $G_{\text{eff}}=\text{constant}$, like the dynamical Chern-Simons gravity \cite{Chern-Simons, AA our first law}, Eq.(\ref{MG GSLT}) reduces to become
\begin{equation}
\big(1+ w_{\text{eff}}\big)\frac{T_m }{T_{\text{A}}}
 \geq
\rho_m A_{\text{A}} G\big( 1+w_m \big)\big(\frac{1}{3} + w_{\text{eff}}\big)  ,
\end{equation}
which appears analogous to Eq.(\ref{GSLT condition 2}) of $\Lambda$CDM.

For the apparent-horizon GSL, these results have matured the pioneering investigations in \cite{2nd law modified gravity} for generic modified gravities and other earlier results in e.g. \cite{EventH GSLT Gauss Bonnet, GSL scalar tensor chameleon} for specific gravity theories by the nonequilibrium revision of the unified first law, selection of the Cai--Kim temperature, dropping of the artificial assumption $T_m=\mathcal{T}^{(+)}_{\text{A}}$, and discovery of the explicit expression for the entropy production  $d_pS^{(\text{A})}$.


\subsection{GSL for the event-horizon system}

For the  event-horizon system, $dS_{\text{E}}+d_pS^{(\text{E})}$ should be directly determined by the nonequilibrium unified first law
Eq.(\ref{dE inside t Upsilon}),
\begin{equation}\label{MG GSL EH 1}
\begin{split}
&T_{\text{E}}\left(dS_{\text{E}}+d_pS^{(\text{E})}\right)\\
=& \delta Q_{(\text{E})}=-
dE_{\text{eff}}^{(\text{E})} \big|_{\Upsilon_{\text{E}0}}=-  \big(A_{\text{E}}\Psi_t+\mathcal{E}_{\text{E}}\big)\\
=& A_{\text{E}}\big(1+w_{\text{eff}}\big)\rho_{\text{eff}} H\Upsilon_{\text{E}} dt
+ V_{\text{E}}\frac{\dot{G}_{\text{eff}}}{G_{\text{eff}}}  \rho_{\text{eff}}  dt  .
\end{split}
\end{equation}
Then Eqs.(\ref{MG EH SSL}) and (\ref{MG GSL EH 1}) yield
\begin{equation}\label{GSL EH system MG}
\begin{split}
\dot{S}_m^{(\text{E})} +\dot{S}_{\text{E}}+\dot{S}_p^{(\text{E})} =\frac{1}{T_m }\rho_m A_{\text{E}}(1+w_m)+ & \\ \frac{1}{T_{\text{E}}}\left(A_{\text{E}} \big(1+w_{\text{eff}}\big)\rho_{\text{eff}} H\Upsilon_{\text{E}}
+ V_{\text{E}}\frac{\dot{G}_{\text{eff}}}{G_{\text{eff}}}   \rho_{\text{eff}}   \right)&.
\end{split}
\end{equation}
Inspired by the validity of the event-horizon GSL for Sec.~\ref{Generalized second law for the event-horizon system} and the standard second law Eq.(\ref{MG EH SSL}), we a priori anticipate
$\dot{S}_m^{(\text{E})} +\dot{S}_{\text{E}}+\dot{S}_p^{(\text{E})}\geq 0$ to hold, which imposes the following viability constraint to modified gravities
\begin{equation}\label{GSL MG EH}
\frac{T_m }{T_{\text{E}}}\left(\big(1+w_{\text{eff}}\big) H
+ \frac{\dot{G}_{\text{eff}}}{G_{\text{eff}}}   \right) \rho_{\text{eff}}
\geq -\rho_m(1+w_m)\Upsilon_{\text{E}}^{-1} .
\end{equation}
Considering that $-1/3\leq w_m\leq 1$, its right hand side  is negative definite, so a sufficient (yet not necessary) condition to validate the GSL is
\begin{equation}
\left(\big(1+w_{\text{eff}}\big) H
+ \frac{\dot{G}_{\text{eff}}}{G_{\text{eff}}}   \right) \rho_{\text{eff}} \geq 0 .
\end{equation}
These results improve the earlier investigations in e.g. \cite{Event H entropy correct II} for the event-horizon GSL in modified gravities.

Note that the discussion in Sec.~\ref{Nonequilibrium Clausius equation On the horizon} is based on the holographic-style gravitational equations and only applies to the apparent-horizon system;
if presuming a Wald entropy $A_{\text{E}}/4G_{\text{eff}}$ and
employing the entropy production to balance all differential terms involving the evolution effect $\dot{G}_{\text{eff}}$, one would obtain
\begin{equation}\label{GSL EH system MG II}
T_{\text{E}}\left(dS_{\text{E}}+d_pS^{(\text{E})}\right) =  \left(  T_{\text{E}} \frac{ 2\pi\Upsilon_{\text{E}}\dot{\Upsilon}_{\text{E}}}{G_{\text{eff}}}
+ V_{\text{E}}\frac{\dot{G}_{\text{eff}}}{G_{\text{eff}}}   \rho_{\text{eff}}   \right)dt ,
\end{equation}
with $d_pS^{(\text{E})}$ specified as
\begin{equation}
d_pS^{(\text{E})} =    \left(T_{\text{E}} A_{\text{E}} \frac{ \dot{G}_{\text{eff}}}{4G_{\text{eff}}^2}
+ V_{\text{E}}\frac{\dot{G}_{\text{eff}}}{G_{\text{eff}}}   \rho_{\text{eff}}   \right)dt .
\end{equation}
Comparison of Eqs.(\ref{MG GSL EH 1}) and (\ref{GSL EH system MG II}) yields the condition
\begin{equation}
T_{\text{E}} \frac{2\pi\Upsilon_{\text{E}}\big(H\Upsilon_{\text{E}}-1\big)}{G_{\text{eff}}} =
A_{\text{E}} \big(1+w_{\text{eff}}\big)\rho_{\text{eff}}  H\Upsilon_{\text{E}} ,
\end{equation}
and thus the whole discussion in Sec.~\ref{Bekenstein-Hawking entropy and Cai--Kim temperature for the event horizon} for $\Lambda$CDM can be parallelly applied to modified gravities with $G\mapsto G_{\text{eff}}$, $\rho_m\mapsto \rho_{\text{eff}}$ and $w_m\mapsto w_{\text{eff}}$, which again implies that the entropy $A/4G_{\text{eff}}$ and the Cai--Kim temperature $1/(2\pi \Upsilon)$ only unconditionally hold on the cosmological event horizon.

%
\subsection{A note on existing methods of GSL}

Existent papers on GSL of modified gravities (in the  traditional positive-heat-in Gibbs equation $T_m d\widehat{S}_m=dE+P_m dV$) usually replace $\rho_m+P_m$ by $\widetilde{\rho}_{\text{(MG)}}+\widetilde{P}_{\text{(MG)}}$ in Eq.(\ref{MG Gibbs 2nd law setup}),
with $\{\widetilde{\rho}_{\text{(MG)}},\widetilde{P}_{\text{(MG)}}\}$ set up in the field equation involving both  Newton's constant $G$ and the dynamic $G_{\text{eff}}$:
\begin{equation}\label{FieldEqnGRForm hybrid}
\begin{split}
R_{\mu\nu}-\frac{1}{2}Rg_{\mu\nu}
 &= 8\pi G  \widetilde{T}_{\mu\nu}^{\text{(eff)}}
  =  8\pi G  \Big(\widetilde{T}_{\mu\nu}^{(m)}+\widetilde{T}_{\mu\nu}^{\text{(MG)}}\Big) ,
\end{split}
\end{equation}
where $\widetilde{T}^{\mu \text{(eff)}}_{\;\;\nu}=\text{diag} \left[-\widetilde{\rho}_{\text{eff}},\widetilde{P}_{\text{eff}},\widetilde{P}_{\text{eff}},\widetilde{P}_{\text{eff}}\right]$,
$\widetilde{\rho}_{\text{eff}}=\widetilde{\rho}_m +\widetilde{\rho}_{\text{(MG)}} $,  $\widetilde{P}_{\text{eff}}=\widetilde{P}_m+\widetilde{P}_{\text{(MG)}}$, and the tilde $\sim$ means that the possibly dynamical aspect of $G_{\text{eff}}$ in Eq.(\ref{FieldEqnGRForm}) has been absorbed into $\widetilde{T}_{\mu\nu}^{\text{(eff)}}$ to formally maintain a constant coupling strength $G$; also note that for these tilded quantities the conservation equation becomes $\dot{\tilde{\rho}}_{\text{eff}}+3H \left(\tilde{\rho}_{\text{eff}}+\tilde{P}_{\text{eff}}  \right)=0$ and $\dot{\rho}_m+3H (\rho_m+P_m  )=0$ under minimal coupling (an energy exchange term between $\rho_m$ and $\widetilde{\rho}_{\text{(MG)}}$ was analyzed for minimal $f(R)$ gravity in \cite{GSLT f R interacting}, which however should be a feature of nonminimal coupling). This way, for the apparent-horizon system with $T_m \dot{\widehat{S}}_m =4\pi \Upsilon_{\text{A}}^2\big(\rho_m+P_m \big) \left(
 \Upsilon_{\text{A}}-H\dot{\Upsilon}_{\text{A}}\right)dt$, one would have the GSL (e.g. \cite{EventH GSLT Gauss Bonnet, GSL scalar tensor chameleon, GSLT f R interacting} for the $F(R,\mathcal{G})$, scalar-tensor-chameleon and interacting $f(R)$ gravities)
\begin{widetext}
\begin{equation}\label{standard GSL old}
\dot{\widehat{S}}_m^{(\text{A})}  +\dot{S}_{\text{A}}  = \frac{1}{T_m}\frac{G}{G_{\text{eff}}} \Bigg(\frac{\dot{\Upsilon}_{\text{A}}}{GH \Upsilon_{\text{A}} }-4\pi \Upsilon_{\text{A}}^2 \big(\widetilde{\rho}_{\text{(MG)}}+\widetilde{P}_{\text{(MG)}}\big)\Bigg)  \Big(\dot{\Upsilon}_{\text{A}} - \Upsilon_{\text{A}} H   \Big) + \frac{ 2\pi\Upsilon_{\text{A}}\dot{\Upsilon}_{\text{A}}}{G_{\text{eff}}} ,
\end{equation}
where $G_{\text{eff}}$ is recognized from the coefficient of $G\widetilde{\rho}_m=G_{\text{eff}}\rho_m$ to utilize the Wald entropy ${S}_{\text{A}}=A_{\text{A}}/4G_{\text{eff}}$. In Eq.(\ref{standard GSL old}) we have incorporated the holographic-style gravitational equations [simply Eqs.(\ref{MG Friedmann Upsilon})-(\ref{MG Friedmann Upsilon IV}) with $G_{\text{eff}}\mapsto G$ and ${\rho}_{\text{eff}}\mapsto\widetilde{\rho}_{\text{eff}}$, ${P}_{\text{eff}}\mapsto\widetilde{P}_{\text{eff}}$] for compactness, as well as the relation
\begin{equation}
\rho_m+P_m=\frac{G}{G_{\text{eff}}} \left( \widetilde{\rho}_m+\widetilde{P}_m\right) .
\end{equation}
However, Eq.(\ref{standard GSL old}) is not self-consistent, not just for the conflicting sign conventions encoded in ${\widehat{S}}_m^{(\text{A})} $ and $S_{\text{A}}$, but also because it uses two different  coupling strength for \{$\widehat{S}_m^{(\text{A})}$, ${S}_{\text{A}}$\}, and fails to capture the extra entropy production $d_pS^{(\text{A})}$ which arises in all modified gravities with nontrivial $G_{\text{eff}}$ \cite{AA our first law, Eling Nonequilibrium}. To overcome these flaws in this popular method, the adjusted Gibbs equation (\ref{Gibbs Positive out}) along with the setups in Eqs.(\ref{FieldEqnGRForm}, \ref{Effective SEM Perfect fluid}) and the holographic-style Eqs.(\ref{MG Friedmann Upsilon})-(\ref{MG Friedmann Upsilon IV}) lead to
\begin{equation}\label{standard GSL new Sm}
\begin{split}
T_m  \dot{S}_m    &=-\frac{1}{G_{\text{eff}}}
\Bigg(\frac{\dot{\Upsilon}_{\text{A}}}{H \Upsilon_{\text{A}} }-4\pi \Upsilon_{\text{A}}^2 G_{\text{eff}} \big(\rho_{\text{(MG)}}+P_{\text{(MG)}}\big)\Bigg)  \Big(\dot{\Upsilon}_{\text{A}} - \Upsilon_{\text{A}} H   \Big)\\
&= -\frac{H\Upsilon_{\text{A}}^5}{G_{\text{eff}}}  \Bigg(\dot H-\frac{k}{a^2}
+4\pi G_{\text{eff}}\big(\rho_{\text{(MG)}}+P_{\text{(MG)}}\big)\Bigg)  \Big( \dot{H}+ H^2 \Big) ,
\end{split}
\end{equation}
which together with Eq.(\ref{MG TdS nonequi}) yields
\begin{equation}\label{standard GSL new}
\dot{S}_m^{(\text{A})} +\dot{S}_{\text{A}}+\dot{S}_p^{(\text{A})} = -\frac{H\Upsilon_{\text{A}}^5}{G_{\text{eff}}}  \Bigg(\dot H-\frac{k}{a^2}
+4\pi G_{\text{eff}}\big(\rho_{\text{(MG)}}+P_{\text{(MG)}}\big)\Bigg)  \Big( \dot{H}+ H^2 \Big)+ \frac{ 2\pi\Upsilon_{\text{A}}\dot{\Upsilon}_{\text{A}}}{G_{\text{eff}}}
+\pi\Upsilon_{\text{A}}^2\frac{\dot{G}_{\text{eff}}}{G_{\text{eff}}^2}  .
\end{equation}
\end{widetext}
 Eq.(\ref{standard GSL new}) improves  Eq.(\ref{standard GSL old}) into a totally self-consistent and more natural method that employs a single
gravitational coupling strength $G_{\text{eff}}$ in accordance with the standard entropy $A_{\text{A}}/4G_{\text{eff}}$. The approach by  Eq.(\ref{standard GSL new}) looks more concentrative on \{$\rho_{\text{(MG)}} ,P_{\text{(MG)}}$\} of the modified-gravity effects; however, it has implicitly ignored the nonexotic character of the cosmic fluid $\rho_m+3P_m\geq 0$, and complicated the mathematical calculations. Hence, in this paper we have chosen to work with Eqs.(\ref{MG Gibbs AppH}, \ref{MG GSL AP 0}) rather than Eqs.(\ref{standard GSL new Sm}, \ref{standard GSL new} ) for the apparent-horizon system, and similarly Eq.(\ref{MG EH SSL}, \ref{GSL EH system MG}) for the event-horizon system.

%
\subsection{Applications to concrete modified gravities}\label{subsection Applications}

The formulation of gravitational thermodynamics in this section applies to all ordinary modified gravities of the form Eq.(\ref{FieldEqnGRForm}). One can just reverse the process and logic in \cite{AA our first law} to see the detailed applications of the first laws for different gravity theories, and in this paper we will focus on the concretization of the second laws, for which we have drawn the following generic conclusions:
\begin{enumerate}
  \item[(1)] $\dot{S}_m^{(\text{E})}>0$ always holds, while  $\dot{S}_m^{(\text{A})}\geq 0$ when $w_{\text{eff}} \leq -1/3$;
  \item[(2)] $\dot{S}_m^{(\text{E})} +\dot{S}_{\text{E}}+\dot{S}_p^{(\text{E})}\geq 0$ should hold with Eq.(\ref{GSL MG EH}) as a validity constraint for modified gravities, while $\dot{S}_m^{(\text{A})} +\dot{S}_{\text{A}}+\dot{S}_p^{(\text{A})}\geq 0$ could conditionally hold only when Eq.(\ref{MG GSLT}) is satisfied.
\end{enumerate}
To concretize these conditions, one just needs to find out the effective gravitational coupling strength $G_{\text{eff}}$, the effective EoS parameter
\begin{equation}
\begin{split}
w_{\text{eff}} &= \frac{(P_{\text{eff}}+\rho_{\text{eff}})-\rho_{\text{eff}}}{\rho_{\text{eff}}}\\
&=-1+\frac{(1+w_m)\rho_m+\left(\rho_{\text{(MG)}}+P_{\text{(MG)}}\right)}{\rho_m+ \rho_{\text{(MG)}}} ,
\end{split}
\end{equation}
the ``modified-gravity energy density'' $\rho_{\text{(MG)}}$, and $\rho_{\text{(MG)}}+P_{\text{(MG)}}$.
\subsubsection{$f(R)$ gravity}
For the FRW Universe governed by the $\mathscr{L}=f(R) +16\pi G \mathscr{L}_m $ gravity \cite{Example fR}, we have $G_{\text{eff}}={G}/{f_R}$ and \cite{AA our first law}
\begin{eqnarray}
\rho_{\text{(MG)}}&=&\frac{1}{8\pi G}  \Big(\frac{1}{2}f_R R-\frac{1}{2}f-3H\dot{f}_R \Big)\label{fR rho P}\\ 
\rho_{\text{(MG)}}&+&P_{\text{(MG)}}
 =\; \frac{1}{8\pi G}  \Big( \ddot{f}_{R}-H\dot{f}_R \Big)\label{fR rho P} \\
w_{\text{eff}}=-&1&+ \frac{8\pi G(1+w_m)\rho_m+ \ddot{f}_{R}-H\dot{f}_R }{8\pi G\rho_m+ \frac{1}{2}f_R R-\frac{1}{2}f-3H\dot{f}_R } .
\end{eqnarray}
The GSL for the event-horizon system requires $f(R)$ gravity to respect the following viability condition
\begin{eqnarray}
&&\frac{T_m }{T_{\text{E}}}\left(\frac{8\pi G(1+w_m)\rho_m+ \ddot{f}_{R}-H\dot{f}_R }{8\pi G\rho_m+ \frac{1}{2}f_R R-\frac{1}{2}f-3H\dot{f}_R } H
-\frac{\dot{f}_R}{f_R}  \right)\times\\
&&\left(8\pi G\rho_m+ \frac{1}{2}f_R R-\frac{1}{2}f-3H\dot{f}_R\right)
\geq -8\pi G\rho_m(1+w_m)\Upsilon_{\text{E}}^{-1}\nonumber,
\end{eqnarray}
while for the apparent-horizon open system, the second law and the GSL respectively hold in the situations
\begin{eqnarray}
\frac{8\pi G(1+w_m)\rho_m+ \ddot{f}_{R}-H\dot{f}_R }{8\pi G\rho_m+ \frac{1}{2}f_R R-\frac{1}{2}f-3H\dot{f}_R }\leq -\frac{2}{3} ,
\end{eqnarray}
\begin{eqnarray}
&&\frac{T_m}{T_{\text{A}}}  \left(f_R \frac{8\pi G(1+w_m)\rho_m+ \ddot{f}_{R}-H\dot{f}_R }{8\pi G\rho_m+ \frac{1}{2}f_R R-\frac{1}{2}f-3H\dot{f}_R}
-\frac{\dot{f}_R}{3H}\right)\geq\\
&&G\rho_m A_{\text{A}} \big( 1+w_m \big)\left(\frac{8\pi G(1+w_m)\rho_m+ \ddot{f}_{R}-H\dot{f}_R }{8\pi G\rho_m+ \frac{1}{2}f_R R-\frac{1}{2}f-3H\dot{f}_R}-\frac{2}{3} \right)\nonumber.
\end{eqnarray}

\subsubsection{Scalar-tensor-chameleon gravity}\label{subsection Scalar tensor chameleon gravity}

For the scalar-tensor-chameleon gravity \cite{GSL scalar tensor chameleon} with the Lagrangian density $\mathscr{L}_{\text{STC}}=F(\phi)R-Z(\phi) \nabla_{\alpha}\phi \nabla^{\alpha}\phi -2U(\phi) +16\pi G E(\phi)\mathscr{L}_m $, we have $G_{\text{eff}}=\frac{E(\phi)}{F(\phi)} G$ and \cite{AA our first law}
\begin{equation}\label{STC rho P}
\rho_{\text{(MG)}} =  \frac{1}{ 8\pi GE }\bigg(-3H\dot F+\frac{1}{2} Z  \dot{\phi}^2+U\bigg)
\end{equation}
\begin{equation}\label{STC weff}
w_{\text{eff}} 
=-1+\frac{8\pi GE (1+w_m)\rho_m+ \ddot F-H\dot F+Z \dot{\phi}^2}{8\pi GE \rho_m  -3H\dot F+\frac{1}{2} Z  \dot{\phi}^2+U} ,
\end{equation}
where in this subsection we temporarily adopt the abbreviations $E\equiv E(\phi)$, $F\equiv F(\phi)$, $U\equiv U(\phi)$ and $Z\equiv Z(\phi)$.
Eq.(\ref{GSL MG EH}) for the GSL of the event--horizon system imposes the constraint
\begin{equation} \label{GSL MG EH STC}
\begin{split}
\hspace{-2mm}&\frac{T_m}{T_{\text{E}}}\left(\frac{8\pi GE (1+w_m)\rho_m+ \ddot F-H\dot F+Z \dot{\phi}^2}{8\pi GE \rho_m  -3H\dot F+\frac{1}{2}Z \dot{\phi}^2+U}H
+ \frac{FE_\phi-EF_\phi}{E F }\dot\phi \right)\\
&\times\left(8\pi G\rho_m-3H\frac{\dot F}{ E }+\frac{Z }{ E }\dot{\phi}^2+\frac{U}{ E }\right)
\geq -8\pi G\rho_m (1+w_m)\Upsilon_{\text{E}}^{-1},
\end{split}
\end{equation}
while $w_{\text{eff}}\leq -\frac{1}{3}$ and the apparent-horizon GSL Eq.(\ref{GSL MG EH}) can be directly realized  with Eq.(\ref{STC weff}) and $\rho_{\text{(MG)}}+P_{\text{(MG)}}= \frac{1}{ 8\pi G E }\left(\ddot F-H\dot F+Z \dot{\phi}^2\right)$.
Moreover, in the specifications $E\mapsto 1$, $F \mapsto \phi$, $Z \mapsto \omega /\phi$, $U\mapsto \frac{1}{2}V$, we recover the generalized Brans-Dicke gravity  \cite{Brans Dicke} with a self-interacting potential, $\mathscr{L}_{\text{GBD}}= \phi R-\frac{\omega }{\phi} \nabla_{\alpha}\phi \nabla^{\alpha}\phi-V(\phi) +16\pi G\mathscr{L}_m$, and Eq.(\ref{GSL MG EH STC}) reduces to become
\begin{eqnarray}
&&\frac{T_m }{T_{\text{E}}}\left(\frac{8\pi G(1+w_m)\rho_m+ \ddot{\phi}-H\dot\phi+\frac{\omega}{\phi}\dot{\phi}^2 }{8\pi G\rho_m -3H\dot\phi+\frac{\omega}{2\phi}\dot{\phi}^2+\frac{1}{2}V}H
-\frac{\dot{\phi}}{\phi} \right)\times\\
&&\left(8\pi G\rho_m -3H\dot\phi+\frac{\omega}{2\phi}\dot{\phi}^2+\frac{V}{2}\right)
\geq -8\pi G\rho_m (1+w_m)\Upsilon_{\text{E}}^{-1}.\nonumber
\end{eqnarray}

\subsubsection{Scalar-tensor-chameleon gravity}\label{subsection Scalar tensor chameleon gravity}

For the scalar-tensor-chameleon gravity \cite{GSL scalar tensor chameleon} with the Lagrangian density $\mathscr{L}_{\text{STC}}=F(\phi)R-Z(\phi) \nabla_{\alpha}\phi \nabla^{\alpha}\phi -2U(\phi) +16\pi G E(\phi)\mathscr{L}_m $ in the Jordan conformal frame, which generalizes the Brans-Dicke gravity,
we have $G_{\text{eff}}=\frac{E(\phi)}{F(\phi)} G$ and \cite{AA our first law}
\begin{equation}\label{STC rho P}
\rho_{\text{(MG)}} =  \frac{1}{ 8\pi GE }\bigg(-3H\dot F+\frac{1}{2} Z  \dot{\phi}^2+U\bigg)
\end{equation}
\begin{equation}\label{STC weff}
w_{\text{eff}} 
=-1+\frac{8\pi GE (1+w_m)\rho_m+ \ddot F-H\dot F+Z \dot{\phi}^2}{8\pi GE \rho_m  -3H\dot F+\frac{1}{2} Z  \dot{\phi}^2+U} ,
\end{equation}
where in this subsection we temporarily adopt the abbreviations $E\equiv E(\phi)$, $F\equiv F(\phi)$, $U\equiv U(\phi)$ and $Z\equiv Z(\phi)$, while $H$
is still the Hubble parameter.
Eq.(\ref{GSL MG EH}) for the GSL of the event--horizon system imposes the constraint
\begin{widetext}
\begin{equation}\label{GSL MG EH STC}
\begin{split}
\frac{T_m}{T_{\text{E}}}\left(\frac{8\pi GE (1+w_m)\rho_m+ \ddot F-H\dot F+Z \dot{\phi}^2}{8\pi GE \rho_m  -3H\dot F+\frac{1}{2}Z \dot{\phi}^2+U}H
+ \frac{FE_\phi-EF_\phi}{E F }\dot\phi \right)
\times\left(8\pi G\rho_m-3H\frac{\dot F}{ E }+ \frac{Z }{ 2E }\dot{\phi}^2+\frac{U}{ E }\right)
\geq -8\pi G\rho_m (1+w_m)\Upsilon_{\text{E}}^{-1},
\end{split}
\end{equation}
while $w_{\text{eff}}\leq -\frac{1}{3}$ and the apparent-horizon GSL Eq.(\ref{GSL MG EH}) can be directly realized  with Eq.(\ref{STC weff}) and $\rho_{\text{(MG)}}+P_{\text{(MG)}}= \frac{1}{ 8\pi G E }\left(\ddot F-H\dot F+Z \dot{\phi}^2\right)$.
Moreover, in the specifications $E\mapsto 1$, $F \mapsto \phi$, $Z \mapsto \omega /\phi$, $U\mapsto \frac{1}{2}V$, we recover the generalized Brans-Dicke gravity  \cite{Brans Dicke} with a self-interacting potential, $\mathscr{L}_{\text{GBD}}= \phi R-\frac{\omega }{\phi} \nabla_{\alpha}\phi \nabla^{\alpha}\phi-V(\phi) +16\pi G\mathscr{L}_m$, and Eq.(\ref{GSL MG EH STC}) reduces to become
\begin{eqnarray}
\frac{T_m }{T_{\text{E}}}\left(\frac{8\pi G(1+w_m)\rho_m+ \ddot{\phi}-H\dot\phi+\frac{\omega}{\phi}\dot{\phi}^2 }{8\pi G\rho_m -3H\dot\phi+\frac{\omega}{2\phi}\dot{\phi}^2+\frac{1}{2}V}H
-\frac{\dot{\phi}}{\phi} \right)\times
\left(8\pi G\rho_m -3H\dot\phi+\frac{\omega}{2\phi}\dot{\phi}^2+\frac{1}{2}V \right)
\geq -8\pi G\rho_m (1+w_m)\Upsilon_{\text{E}}^{-1}\,.
\end{eqnarray}

\subsubsection{Quadratic gravity}
For the quadratic gravity $\mathscr{L}_{\text{QG}}= R+ a R^2 +b R_{\mu\nu}R^{\mu\nu}  + 16\pi G \mathscr{L}_m $ whose Lagrangian density is an effective linear superposition of the quadratic independent Riemannian invariants \cite{Example Quardratic gravity second paper, AA Tian-Booth Paper}, with $\{a,b\}$ being constants, we have
$G_{\text{eff}}=\frac{G}{1+2aR}$ and \cite{AA our first law}
\begin{equation}
\rho_{\text{(MG)}}=\frac{1}{8\pi G} \Bigg(\frac{a}{2}R^2- \frac{b}{2}R_c^2 +\frac{b}{2} \ddot R-\big(4a+b\big)  H\dot R
+4b  R^t_{\;\;\alpha t\beta} +2b \Box R_t^{\;\;t} \Bigg),
\end{equation}
\begin{equation}\label{QG weff}
\begin{split}
w_{\text{eff}} 
&=-1+\frac{8\pi G(1+w_m)\rho_m+ \big(2a+b\big) \ddot{R}-\frac{b}{2}  H\dot{R}
+4b (R^t_{\;\;\alpha t\beta}-R^r_{\;\;\alpha r\beta})R^{\alpha\beta} +2b \Box \big(R_t^{\;\; t}-R_r^{\;\;r}\big)}{8\pi G\rho_m+  \frac{a}{2}R^2- \frac{b}{2}R_c^2 +\frac{b}{2} \ddot R-\big(4a+b\big)  H\dot R
+4b  R^t_{\;\;\alpha t\beta} +2b \Box R_t^{\;\;t}} ,
\end{split}
\end{equation}
where $R_c^2\coloneqq R_{\mu\nu}R^{\mu\nu}$, $\Box=g^{\mu\nu}\nabla_\mu\nabla_\nu$, and we have used the compact geometric notations \cite{AA our first law}. Hence, GSL of the event-horizon system requires
\begin{equation}
\begin{split}
\frac{T_m }{T_{\text{E}}}\left(\frac{8\pi G(1+w_m)\rho_m+ \big(2a+b\big) \ddot{R}-\frac{b}{2}  H\dot{R}
+4b (R^t_{\;\;\alpha t\beta}-R^r_{\;\;\alpha r\beta})R^{\alpha\beta} +2b \Box \big(R_t^{\;\; t}-R_r^{\;\;r}\big)}{8\pi G\rho_m+  \frac{a}{2}R^2- \frac{b}{2}R_c^2 +\frac{b}{2} \ddot R-\big(4a+b\big)  H\dot R
+4b  R^t_{\;\;\alpha t\beta} +2b \Box R_t^{\;\;t}} H
-\frac{2a \dot{R}}{1+2aR}\right)\\
\times\left(8\pi G\rho_m+\frac{a}{2}R^2- \frac{b}{2}R_c^2 +\frac{b}{2} \ddot R-\big(4a+b\big)  H\dot R+4b  R^t_{\;\;\alpha t\beta} +2b \Box R_t^{\;\;t}\right) \geq -8\pi G\rho_m (1+w_m)\Upsilon_{\text{E}}^{-1} ,
\end{split}
\end{equation}
while $w_{\text{eff}}\leq -\frac{1}{3}$ and Eq.(\ref{GSL MG EH})
 can be directly concretized with Eq.(\ref{QG weff}) and
\begin{equation}
\begin{split}
\rho_{\text{(MG)}}+P_{\text{(MG)}} =
\frac{1}{8\pi G} \Bigg( \big(2a+b\big) \ddot{R}-\frac{b}{2}  H\dot{R}
+4b (R^t_{\;\;\alpha t\beta}-R^r_{\;\;\alpha r\beta})R^{\alpha\beta} +2b \Box \big(R_t^{\;\; t}-R_r^{\;\;r}\big) \Bigg) .
\end{split}
\end{equation}
\end{widetext}

\subsubsection{Chern-Simons gravity}
Finally let's analyze the dynamical Chern-Simons gravity $\mathscr{L}_{\text{CS}}= R+\frac{a \vartheta}{\sqrt{-g}} {}^*\widehat{RR}-b \nabla_\mu\vartheta \nabla^\mu\vartheta-V(\vartheta)+16\pi G\mathscr{L}
_m$ \cite{Chern-Simons} which has a constant gravitational coupling strength $G_{\text{eff}}=G$, where ${}^*\widehat{RR}=  {}^{*}R_{\alpha\beta\gamma\delta} R^{\alpha\beta\gamma\delta}$ denotes the Chern-Pontryagin invariant and $\{a,b\}$ are constants. We have  \cite{AA our first law}
\begin{eqnarray}
\rho_{\text{(MG)}}&=& \frac{1}{16\pi G} \left(b \dot{\vartheta}^2+  V(\vartheta)\right)\label{CS rho P}\\
w_{\text{eff}} =-1&+&\frac{8\pi G\rho_m(1+w_m)+ b \dot{\vartheta}^2}{8\pi G\rho_m+ \frac{1}{2}b \dot{\vartheta}^2+  \frac{1}{2}V(\vartheta)} ,
\end{eqnarray}
and thus Eq.(\ref{GSL MG EH}) leads to the viability condition
\begin{equation}\label{GSL MG EH DCS 0}
\frac{T_m }{T_{\text{E}}}
\left(8\pi G\rho_m(1+w_m)+ b \dot{\vartheta}^2\right)\geq -8\pi G\rho_m (1+w_m)\frac{\Upsilon_{\text{H}}}{\Upsilon_{\text{E}}} ,
\end{equation}
which, for $\dot{\vartheta}\neq 0$, yields a constraint for $b$,
\begin{equation}\label{GSL MG EH DCS}
 b \geq
-8\pi G\rho_m (1+w_m)\left(\frac{\Upsilon_{\text{H}}}{\Upsilon_{\text{E}}}\frac{T_{\text{E}}}{T_m }+1\right)\dot{\vartheta}^{-2} .
\end{equation}
For the FRW cosmology, ${}^*\widehat{RR}$ makes no contribution to the gravitational equations,  so $\mathscr{L}_{\text{CS}}$ effectively acts as $\mathscr{L}= R-b \nabla_\mu\vartheta \nabla^\mu\vartheta-V(\vartheta)+16\pi G\mathscr{L}_m$, which formally resembles the scalarial dark energy \cite{DE Quintessence, DE phantom}.  On the other hand, note that although Eqs.(\ref{GSL MG EH DCS 0}) and (\ref{GSL MG EH DCS}) are always satisfied for $b>0$, which corresponds to a canonical kinetic $\vartheta$-field that is quintessence-like ($\mathscr{L}= - \frac{1}{2} \nabla_\mu\phi \nabla^\mu\phi-V(\phi)$), $\vartheta$ is allowed to be slightly phantom-like ($\mathscr{L}= \frac{1}{2} \nabla_\mu\phi \nabla^\mu\phi-V(\phi)$) for some $b<0$ by Eq.(\ref{GSL MG EH DCS}). Hence, Eq.(\ref{GSL MG EH DCS}) does not coincide with the situation of $\Lambda$CDM in Sec.~\ref{Generalized second law for the event-horizon system}, where $\dot{S}_m^{\text{(E)}} +\dot{S}_{\text{E}}\geq0$ holds if and only if $w_m\geq-1$.


\section{Conclusions and discussion}

In this paper the thermodynamic implications of the holographic-style dynamical equations for the FRW Universe have been studied. We started from the $\Lambda$CDM model of GR  to clearly build the whole framework of gravitational thermodynamics, and eventually extended it to modified gravities. A great advantage of our formulation is all constraints are expressed by the EoS parameters.

The holographic-style gravitational equations govern both the apparent-horizon dynamics and  the cosmic spatial expansion. We have shown how they imply Hayward's unified first law of equilibrium thermodynamics $dE=A\bm\psi+WdV$ \cite{Hayward Unified first law} and the isochoric-process Cai--Kim Clausius equation $T_{\text{A}}dS_{\text{A}}=\delta Q_{\text{A}}=-A_{\text{A}}\bm\psi_t$ \cite{Cai I, Cai IV, Temperature tuneling}.
The derivations of the Clausius equation in Sec.~\ref{GR Clausius equation On the horizon} actually involves a long standing confusion regarding the setup of the apparent-horizon temperature, and extensive comparisons in  Sec.~\ref{Solution to the horizon-temperature confusion} have led to the argument that
 the Cai--Kim $T_{\text{A}}=1/(2\pi \Upsilon_{\text{A}})$
is more appropriate than the Hayward $\mathcal{T}_{\text{A}}={\kappaup}/{2\pi}$ and its partial absolute value $\mathcal{T}_{\text{A}}^{(+)}$. Meanwhile, we have also introduced the ``zero temperature divide'' $w_m=1/3$ for  $\mathcal{T}_{\text{A}}={\kappaup}/{2\pi}$, and proved the signs of both temperatures are independent of the inner or outer trappedness of the apparent horizon.

The ``positive heat out''  sign convention for the heat transfer and the horizon entropy has been decoded from $T_{\text{A}}dS_{\text{A}}=-dE$, provided that the third law of thermodynamics holds with a positive $T_{\text{A}}$. With the horizon temperature and entropy clarified, the cosmic entropy evolution has been investigated. We have adjusted the traditional matter entropy and Gibbs equation  into $dE_m=-T_mdS_m-P_mdV_{\text{A}}$ in accordance with the positive heat  out convention of the horizon entropy. It turns out that the cosmic entropy is well behaved, specially for the event-horizon system, where both the second law and the GSL hold for nonexotic matter  ($-1\leq w_m\leq 1$). Also, we have clarified that the regions \{$\Upsilon\leq \Upsilon_{\text{A}}$, $\Upsilon\leq \Upsilon_{\text{E}}$\} enveloped by the apparent  and  even horizons are  simple open thermodynamic  systems\footnote{Even the philosophical ``whole Universe'' would be an open system if there were matter creations which would cause irreversible extra entropy production, and one typical mechanism triggering this effect is nonminimal curvature-matter coupling  \cite{matter creation 2}.} so that  one should not \emph{a priori} expect the validity of nondecreasing entropy, and abandoned the local equilibrium assumptions restricting the interior and the boundary temperatures.

Finally we have generalized  the whole formulations from the $\Lambda$CDM model  to
ordinary modified gravities  whose field equations have been intrinsically compactified into the GR form  $R_{\mu\nu}-Rg_{\mu\nu}/2=8\pi G_{\text{eff}}  T_{\mu\nu}^{\text{(eff)}}$. To our particular interest, we found that inside the apparent horizon the second law $\dot{S}_m\geq 0$ nontrivially holds if  $w_{\text{eff}} \leq -1/3$, while inside the event horizon  $\dot{S}_m\geq 0$ always validates regardless of the gravity theories in use. These generic conclusions have been concretized in  $f(R)$, scalar-tensor-chameleon, quadratic and dynamical Chern-Simons gravities.

Note that the volume $V$ and surface area $A$ used throughout this paper are interpreted as flat-space quantities in \cite{Bak Cosmic holography}. However, $\Upsilon$ and $A$ are the proper radius and area for the standard sphere $\mathbb{S}^2$ in the $2+2$ (rather than $3+1$) decomposition $ds^2=h_{\alpha\beta}dx^\alpha dx^\beta+\mathbb{S}^2$ of Eq.(\ref{FRW metric I}), while the role of $V$ as a proper quantity is still not clear.

There are still some interesting problems arising in this paper and yet unsolved.
For example, the discussion in Sec.~\ref{Bekenstein-Hawking entropy and Cai--Kim temperature for the event horizon} further raises the question that, what is the temperature $T_{\text{E}}$ for the event horizon? Note that if $T_{\text{E}}\neq T_{\text{A}}$, there would be a spontaneous heat flow between $\Upsilon_{\text{A}}$ and $\Upsilon_{\text{E}}$ -- would it affect the cosmic expansion? On the other hand, it is not clear whether or not the apparent and the event horizons could be heated by the absolute Hubble energy flow and consequently $T_{\text{E}}=T_m$ and $T_{\text{A}}=T_m$: this would avoid the temperature gradient between  $\Upsilon_{\text{A}}$ and $\Upsilon_{\text{E}}$, but throughout this paper we have not yet seen any evidence for $T_{\text{A}}$ to be heated into $T_m$.

Moreover, besides the traditional GSLs, the ``cosmic
holographic principle'' in \cite{Bak Cosmic holography} which argues that the physical entropy $\widehat{S}_m^{\text{(A)}}$ inside the apparent horizon $\Upsilon_{\text{A}}$ could never exceed the apparent-horizon entropy $S_{\text{A}}$, is also problematic in comparing  $\widehat{S}_m^{\text{(A)}}$ with $S_{\text{A}}$ -- this principle should be restudied in the unified positive-heat-out sign convention. Moreover, is $\Upsilon_{\text{A}}$ the only hologram membrane  for the FRW Universe? Can the relative evolution equation (\ref{GR synchronization 3}) be used in astrophysical and cosmological simulations? Also, how would the cosmic entropy evolve in a contracting Universe? We hope to find out the answers in  prospective studies.


\section*{Acknowledgement}

\noindent This work was financially supported by the Natural Sciences and Engineering Research Council of Canada.

\appendix

\section{The minimum set of state functions  and reversibility}\label{The minimum set of state functions  and reversibility}

Eqs.(\ref{GR Hawking-Bekenstein entropy}) and (\ref{T}) clearly indicate that just like ordinary thermodynamics, the geometrically defined horizon temperature $T_{\text{A}}$ and horizon entropy
$S_{\text{A}}$ remain as \textit{state functions}, which are independent of thermodynamic  processes that indeed correspond to
the details   of  cosmic  expansion  $\dot{a}(t)$ and  the apparent-horizon evolution $\dot{\Upsilon}_{\text{A}}$. Just like the regular temperatures of thermodynamic systems, the Cai--Kim  $T_{\text{A}}$ remains as an intensive property with $T_{\text{A}}=T_{\text{A}}(t)=1/2\pi \Upsilon_{\text{A}}(t)$; one should not treat it as an extensive property by $T_{\text{A}}=T_{\text{A}}(V_{\text{A}})=1/(2\pi \sqrt[3]{\frac{3}{4\pi}V_{\text{A}}})$.
Some other state functions involved here include the apparent-horizon radius $\Upsilon_{\text{A}}$, the energy density $\rho_m(t)$, the pressure $P_m(t)$  and thus the EoS parameter $w_m=\rho_m/P_m$.
These state quantities are not totally independent as they are related with one another by the Friedmann equations (\ref{GR Friedmann eqn 1st}), the holographic-style dynamical equation (\ref{GR Friedmann Upsilon}),  and the thermodynamic relations in Secs.~\ref{GR Unified first law of thermodynamics} and ~\ref{GR Clausius equation On the horizon}. Here we select the following quantities to comprise a minimum set of \emph{independent} state functions  for Secs.~\ref{section FRW setups} and ~\ref{GR first laws Thermodynamics}:
\begin{equation}\label{GR minimum set}
\text{Minimum set}\,=\,\Big\{\,\rho_m\,,w_m\,, T_{\text{A}}   \Big\}\,.
\end{equation}
Based on this  set, the product of $\rho_m$ and $w_m$ yields the pressure $P_m$. Through Eq.(\ref{GR Arho}) $\rho_m$ recovers the horizon area $A_{\text{A}}$ and thus determines the entropy  $S_{\text{A}}$. Treating $T_{\text{A}}$ as an intensive property, we do not take the approach from Eq.(\ref{GR Friedmann Upsilon}) or Eq.(\ref{GR Arho}) for the recovery $\rho_m\to A_{\text{A}}\to \Upsilon_{\text{A}}\to T_{\text{A}}$, and instead
let $T_{\text{A}}$ enter the minimum set directly as the Cai--Kim temperature ansatz. Similarly, for modified gravities with the dynamical equations (\ref{MG Friedmann Upsilon})-(\ref{MG Friedmann Upsilon IV}), we choose the minimum set to be $\left\{\rho_{\text{eff}},w_{\text{eff}}, G_{\text{eff}}, T_{\text{A}}\right\}$.

The fact that Eq.(\ref{GR TdS nonequi}) is the Clausius equation for (quasi)equilibrium or reversible thermodynamic processes without extra entropy production raises the question that, what does reversibility mean from the perspective of cosmic and apparent-horizon dynamics?
From the explicit expression of the heat transfer $\delta Q_{\text{A}}=T_{\text{A}}dS_{\text{A}}=A_{\text{A}}\big(1+w_m\big)\rho_mH\Upsilon_{\text{A}}dt$ where
the state quantity $T_{\text{A}}dS_{\text{A}}$ is balanced by the process quantity $\delta Q_{\text{A}}$,
 we naturally identify $H$ as a \emph{process quantity}; moreover,  if one reverses the initial and final states of  $T_{\text{A}}dS_{\text{A}}$,  the state quantities  $\{\rho_m(t)\,,w_m\,,\Upsilon_{\text{A}}\,,A_{\text{A}}\}$  can be automatically reversed. Hence, by reversibility we mean
an imaginary negation  $-H$ of the Hubble parameter that results in a spatial contraction process which directly evolves the Universe from a later state back to the earlier state of  $T_{\text{A}}dS_{\text{A}}$ without reversing the time arrow and causing energy dissipation.

\cite{Irreversible thermodynamics} suggests that since the energy-matter crossing  the apparent horizon for the (accelerated) expanding Universe will not come back in the future,  it should cause extra entropy production, and \cite{Irreversible thermodynamics} further introduced  the entropy flow vector and the entropy production density for it.   In fact, the reversibility of $T_{\text{A}}dS_{\text{A}}=\delta Q_{\text{A}}$ simply allows for such a possibility in principle, rather than the realistic occurrence of the reverse process, so we believe that the entropy-production treatment in \cite{Irreversible thermodynamics} is inappropriate. As shown in Sec.~\ref{Nonequilibrium Clausius equation On the horizon}, irreversibility and entropy production is a common feature for such (minimally coupled) modified gravities with a \textit{nontrivial}  effective gravitational coupling strength  ($G_{\text{eff}}\neq\text{constant}$) when their field equations are cast into the GR form $R_{\mu\nu}-Rg_{\mu\nu}/2=8\pi G_{\text{eff}}  T_{\mu\nu}^{\text{(eff)}}$, and the time evolution of  $G_{\text{eff}}$ causes irreversible energy dissipation and constitutes the only source of  entropy production.

\end{document}